\documentclass[a4paper,onecolumn,11pt,accepted=2026-10-13]{quantumarticle}

\pdfoutput=1
\usepackage{hyperref}
\hypersetup{hypertexnames=false}
\usepackage{float}
\usepackage{minitoc}
\usepackage[toc,page,header]{appendix}
\usepackage{physics}
\usepackage{xfrac}
\usepackage{graphicx}
\usepackage{subfigure}
\usepackage{mathdots}
\usepackage{amsfonts,amssymb,amsmath}
\usepackage{mathtools}
\usepackage{bm}
\usepackage{mathrsfs}
\usepackage{amsthm}
\usepackage{csquotes}
\MakeOuterQuote{"}
\usepackage{epstopdf}
\usepackage{tikz}
\usetikzlibrary{arrows.meta}
\usetikzlibrary{positioning}
\usetikzlibrary{calc}
\usetikzlibrary{decorations.pathreplacing}
\usepackage{paralist}
\usepackage{diagbox}
\usepackage[inline]{enumitem}
\usepackage{qcircuit}
\usepackage{dsfont}
\usepackage{algorithm}
\usepackage{algpseudocode}
\usepackage{color,soul}
\usepackage[numbers,sort&compress]{natbib}

\def\identity{\leavevmode\hbox{\small1\kern-3.8pt\normalsize1}}

\newtheorem{theorem}{Theorem}

\newtheorem{lemma}{Lemma}
\newtheorem{proposition}{Proposition}

\newcommand{\caH}{\mathcal{H}}

\newcommand{\caL}{\mathcal{L}}

\newcommand{\caO}{\mathcal{O}}

\newcommand{\rmi}{\mathrm{i}}

\newcommand{\rmA}{\mathrm{A}}

\renewcommand{\epsilon}{\varepsilon}

\def\eqref#1{\textup{(\ref{#1})}}
\newcommand{\eref}[1]{Eq.~\textup{(\ref{#1})}}

\newcommand{\erefrange}[2]{Eqs.~\textup{(\ref{#1})--(\ref{#2})}}

\newcommand{\lref}[1]{Lemma~\ref{#1}}

\newcommand{\thref}[1]{Theorem~\ref{#1}}

\newcommand{\eqsref}[2]{Eqs.~(\ref{#1}) and (\ref{#2})}

\newcommand{\sref}[1]{Sec.~\ref{#1}}

\newcommand{\fref}[1]{Fig.~\ref{#1}}

\newcommand{\aref}[1]{Appendix~\ref{#1}}

\def\<{\langle}  
\def\>{\rangle}  

\newcommand{\otoc}{\mathrm{OTOC}}
\newcommand{\bbR}{\mathbb{R}}
\newcommand{\rmS}{\mathrm{S}}

\begin{document}
\title{A Symmetry-Enabled Direct Quantum Protocol for Many-Body Green’s Functions}

\author{Changhao Yi}
\email{cyi@shu.edu.cn}
\affiliation{Department of Physics, Shanghai University, Shanghai, China}
\author{Cunlu Zhou}
\email{cunlu.zhou@usherbrooke.ca}
\affiliation{Department of Computer Science, Universit\'e de Sherbrooke, QC, Canada}
\affiliation{Institut quantique, Universit\'e de Sherbrooke, QC, Canada}
\affiliation{Mila – Québec AI Institute, Montréal, QC, Canada}

\begin{abstract}
We present a symmetry-enabled direct quantum protocol for computing many-body Green’s functions, a central tool for studying strongly correlated quantum systems. Our protocol relies only on native time evolution and straightforward measurements available on current hardware platforms. By exploiting parity symmetry, a property satisfied by a broad class of Hamiltonians in condensed matter physics and quantum chemistry, including the Fermi--Hubbard and Heisenberg models, we introduce a tailored quench spectroscopy scheme that recovers both the real and imaginary parts of two-point time correlators, from which Green’s functions can be reconstructed via efficient classical signal processing. We further develop a tailored parity-preserving quantum sampler that prepares parity-resolved (symmetric and antisymmetric) thermal states, enabling finite-temperature extensions within the same framework. Finally, we show that the same symmetry-based measurement primitive extends naturally to out-of-time-ordered correlators. Our results provide a practical route to estimating symmetry-resolved dynamical correlators on near-term and early fault-tolerant quantum hardware.
\end{abstract}

\maketitle

\section{Introduction}

Many-body Green's functions are essential tools for probing strongly interacting many-body systems \cite{altland2010condensed}. They compactly describe the propagation of particles and excitations without explicit wave-function representations, and they encode key information about the density of states and band structure, which underpin phenomena such as superconductivity, magnetism, and phase transitions. Experimentally, spectroscopic techniques can be used to extract Green’s-function information: for example, angle-resolved photoemission spectroscopy (ARPES) \cite{zhang2022angle} measures the momentum-resolved spectral function associated with single-particle Green’s functions, while resonant inelastic X-ray scattering (RIXS) \cite{ament2011resonant} probes collective excitations encoded in higher-order correlation functions. On the computational side, classical approaches face severe limitations when calculating these dynamical quantities. For example, exact diagonalization \cite{viswanath_mueller_1994} is restricted to small clusters due to the exponential growth of the Hilbert space. While quantum Monte Carlo methods \cite{foulkes2001quantum} scale better for static properties, they struggle with the dynamical sign problem and the ill-conditioned analytic continuation required to extract real-time Green’s functions from imaginary time. Similarly, tensor network states \cite{vidal2004efficient}, though powerful in 1D, are limited by the rapid growth of entanglement entropy during non-equilibrium time evolution, which requires an exponentially growing bond dimension to maintain accuracy.

At a fundamental level, Green’s functions are time-ordered correlation functions. Quantum computing offers a promising route to access them by enabling coherent real-time evolution and the measurement of dynamical correlators, potentially bypassing several bottlenecks faced by classical simulations. Many prominent quantum algorithms have been proposed, such as methods based on phase estimation \cite{bauer2016hybrid,loaiza2024nonlinear,sun2025probing,cruz2025quantum,baez2020dynamical,cruz2023superresolution} and variational approaches \cite{chen2021variational,huang2022variational,dhawan2024quantum}. However, such methods often rely on ancilla-assisted interference circuits and controlled time evolution to extract complex-valued correlators, which can be prohibitively resource-intensive on near-term hardware. In particular, controlled implementations of large unitaries substantially increase circuit depth and error rates, while also placing stringent demands on gate fidelity and coherence. These challenges are especially acute for current noisy intermediate-scale quantum (NISQ) devices \cite{preskill2018quantum}, and remain nontrivial even in the early fault-tolerant regime \cite{katabarwa2024early}.

To circumvent these overheads, a complementary class of quantum algorithms has recently gained traction by avoiding controlled unitaries and, in many cases, ancilla qubits. For instance, \emph{quench spectroscopy} (QS) \cite{knap2013probing,menu18quench,despres2019correlation,yang2024resource,sun2025probing,villa2019unraveling}, a widely used protocol inspired by Ramsey interferometry, can be used to probe dynamical response functions. However, early studies such as \cite{villa2019unraveling} focused on fermionic correlation functions under translation symmetry, which restricts the accessible information to the imaginary part of time correlators. Closely related, the \emph{linear-response framework} (LRF) \cite{Kubo1966,kokcu2024linear, Piccinelli2026circuit, bishop2025quantum} exploits the connection between Green's functions and response theory by introducing a weak time-dependent perturbation, $H(t)=H+\epsilon A(t)$, and extracting correlators from the leading-order contribution to measured expectation values such as $\langle A(t)\rangle$. While conceptually simple, this approach typically requires simulating time-dependent Hamiltonians, and the recovered correlators are generally approximate due to higher-order corrections (except in \cite{bishop2025quantum}). Another route is \emph{quantum imaginary time evolution} (QITE) \cite{motta2020determining,yang2024phase,wang2025computing}, which implements normalized imaginary-time evolution (e.g., via Trotter--Suzuki decompositions) and introduces an imaginary-time shift $t\to t+i\tau$. Multi-time correlators can then be obtained from derivatives with respect to $\tau$, but the accuracy is controlled by the asymptotic limit $\tau\to 0$. The \emph{direct Krylov space} (DKS) method \cite{GreeneDiniz2024quantumcomputed,irmejs2025approximating} instead computes Green's functions recursively using analytically derived coefficients estimated from states, observables, and the Hamiltonian, thereby avoiding explicit quantum dynamics. However, this advantage comes at the cost of a high sample complexity that grows with the problem size. Moreover, subspace methods such as DKS can be sensitive to noise and estimation errors, particularly as the recursion depth increases \cite{irmejs2025approximating, atheory22epperly}. Finally, the \emph{direct measurement} scheme \cite{mitarai2019methodology} leverages mid-circuit measurements (MCMs) to access the real part of time correlators without assumptions on the initial state. Despite its generality and simplicity, MCM-based protocols remain challenging on current hardware due to slow and noisy MCMs and crosstalk that can disturb nearby qubits.

\begin{figure*}
    \centering
    \includegraphics[width=0.9\linewidth]{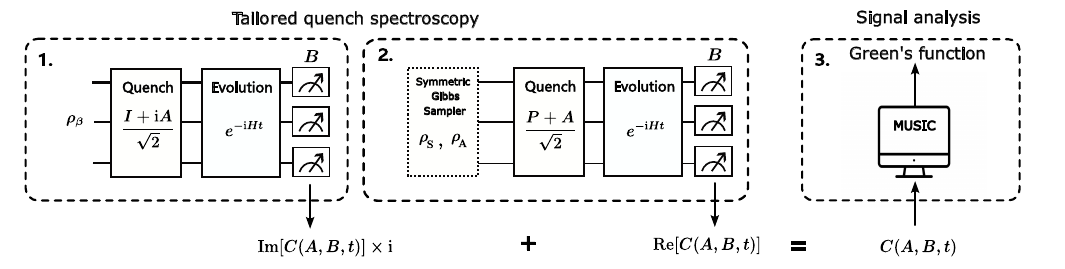}
    \caption{Direct quantum protocol for computing many-body Green’s functions via tailored quench spectroscopy. For an initial thermal state $\rho_\beta$ (with $\beta\to\infty$ corresponding to the zero-temperature limit, so that the thermal state becomes the ground state), the imaginary and real parts of the two-point time correlator $C(A,B,t)$ are obtained by applying the tailored quench operators $(I+\rmi A)/\sqrt{2}$ and $(P+A)/\sqrt{2}$, respectively, followed by time evolution under \(e^{-\rmi H t}\) and measurements in the eigenbasis of the observable $B$. For a thermal initial state, the symmetric and antisymmetric thermal states $\rho_\mathrm{S}$ and $\rho_\mathrm{A}$ are prepared using a tailored parity-preserving quantum sampler (step 2). For a ground-state initial condition, this thermal-state preparation step is not needed. The full Green’s function is then reconstructed via classical signal-processing routines (e.g., MUltiple SIgnal Classification (MUSIC)).}
    \label{fig:protocol}
\end{figure*}

In this work, we introduce an ancilla-free and noise-robust direct quantum protocol for estimating both the real and imaginary parts of two-point time correlators by exploiting \emph{parity symmetry} \cite{altland2010condensed}, from which Green’s functions can be reconstructed via efficient classical signal analysis. We refer to this approach as \emph{tailored quench spectroscopy} (TQS). Specifically, we consider Hamiltonians that commute with a parity operator $P$, while the observables of interest $A$ and $B$ anti-commute with $P$. This condition is satisfied by a broad class of models central to condensed matter physics and quantum chemistry, including the Fermi--Hubbard and Heisenberg models. An overview of the protocol is provided in \fref{fig:protocol}.

By designing tailored quench operators, we extend the standard QS method \cite{villa2019unraveling} to access bosonic correlation functions, enabling direct evaluation of the real part of time correlators. Compared to other ancilla-free protocols, such as LRF and QITE, our TQS approach offers several advantages:
\begin{inparaenum}[i)]
\item it requires only time-independent Hamiltonian evolution, thereby avoiding the simulation of time-dependent dynamics and reducing experimental complexity;
\item it enables exact, non-perturbative recovery of correlators, substantially mitigating systematic errors; and
\item unlike imaginary-time methods, which perturb the state via $e^{-\tau O}$ and typically require additional overhead (e.g., Trotterization or dissipative evolution), it relies on considerably simpler operations.
\end{inparaenum}

In developing TQS protocols for finite-temperature Green's functions, we also introduce a tailored \emph{parity-preserving quantum sampler} for preparing symmetric and antisymmetric thermal states, which may be of independent interest.

The remainder of the paper is organized as follows. In \sref{sec:background}, we review the relationship between two-point time correlators and many-body Green's functions, parity symmetry, and the quench spectroscopy approach. In \sref{sec:mainprotocol}, we present the symmetry-enabled TQS protocol in detail. We generalize the method to out-of-time-ordered correlators (OTOCs) in \sref{sec:otoc}. Complexity and error analyses are provided in \sref{sec:analysis}, and numerical simulations are reported in \sref{sec:numeric}. We conclude in \sref{sec:conclude} with a discussion of future directions.

\section{Background}
\label{sec:background}
\subsection{Many-body Green's functions and two-point time correlators}

We begin by clarifying the relationship between two-point time correlators and the Lehmann representation of many-body Green's functions. Consider a Hamiltonian $H$ with spectral decomposition $H=\sum_{n=0}^{d-1}E_n\,|n\rangle\langle n|$, and let $O_a$ and $O_b$ be two operators. In many-body physics, the two-point correlator is
\begin{equation}
    C(O_a,O_b,t_1,t_2)\coloneqq\mathrm{tr}\!\big[\rho\, O_a(t_1)\,O_b(t_2)\big],
\end{equation}
where $O(t)\coloneqq e^{\rmi Ht}Oe^{-\rmi Ht}$. If the initial state is stationary with respect to $H$, e.g., a Gibbs state $\rho = \rho_\beta \coloneqq e^{-\beta H}/\mathrm{tr}(e^{-\beta H})$ that commutes with $H$, then the correlator depends only on the time difference and can be written as
\begin{equation}\label{eq:correlator}
     C(O_a,O_b,t)=\mathrm{tr}\!\big[\rho\, O_a\, O_b(t)\big].
\end{equation}
Using the spectral decomposition of $H$, we obtain the expansion
\begin{equation}
    C(O_a,O_b,t)=\sum_{m,n}\langle n|O_b|m\rangle\langle m|\rho O_a|n\rangle\,e^{\rmi(E_n - E_m)t}.
\end{equation}
We define the Green's function by a one-sided Fourier (Laplace) transform,
\begin{equation}\label{eq:laplace}
G(z)=-\rmi\int_{0}^{\infty}\!dt\, e^{\rmi zt}\, C(O_a,O_b,-t),
\quad \Im z>0,
\end{equation}
which yields the Lehmann representation \cite{odashima2016pedagogical}
\begin{equation}\label{eq:greenfun}
G(z)
=\sum_{m,n}
\frac{\langle n|O_b|m\rangle\,\langle m|\rho O_a|n\rangle}
{z-(E_n-E_m)},
\quad \Im z>0.
\end{equation}
Green’s functions are commonly expressed either in the time domain, through the correlator $C$, or in the frequency domain, through \eref{eq:greenfun}; in this work we focus on the Lehmann representation. See \aref{app:background-green} for additional background.

In fermionic systems, $O_a$ and $O_b$ are typically linear combinations of fermionic annihilation and creation operators $\{c_i,c_i^\dag\}_i$. For example, if $O_a=\sum_i a_i c_i$ and $O_b=\sum_{i'} b_{i'} c_{i'}^\dag$, then by linearity
\begin{equation}
    C(O_a,O_b,t)=\sum_{i,i'} a_i b_{i'}\, C(c_i,c_{i'}^\dag,t).
\end{equation}
Thus, it suffices to estimate correlators of the form $C(c_i,c_{i'}^\dag,t)$. Under the Jordan--Wigner transformation (see \aref{app:jordan-wigner}), each fermionic operator can be written as a linear combination of two Pauli strings (equivalently, Majorana operators), i.e., $c_i=(A+\rmi B)/2$ with Pauli operators $A$ and $B$ that anticommute. Writing $c_i=(A_0+\rmi B_0)/2$ and $c_{i'}=(A_1+\rmi B_1)/2$, we obtain
\begin{equation}
\begin{aligned}
    \frac{1}{2}C(A_0 + \rmi B_0, A_1 - \rmi B_1,t)
    &= \frac{1}{2}C(A_0,A_1,t)+\frac{1}{2}C(B_0,B_1,t)\\
    &\quad -\frac{\rmi}{2}C(A_0,B_1,t)+\frac{\rmi}{2} C(B_0,A_1,t).
\end{aligned}
\end{equation}
Consequently, the problem reduces to estimating general two-point correlators of the form $C(A,B,t)$, which will be the primary focus of the following sections.

\subsection{Parity symmetry}

If a unitary Hermitian operator $P$ satisfies $[H,P]=0$, then we say that $H$ has a parity symmetry and refer to $P$ as a parity operator. This condition is common in many lattice models. For example, the Heisenberg (XXZ) chain
\begin{align}\label{eq:XXZ}
    H_{\mathrm{XXZ}} &= J_X\left(\sum_{n=0}^{N-2}X_n X_{n+1} + Y_n Y_{n+1}\right) + J_Z \sum_{n=0}^{N-2}Z_n Z_{n+1} + h\sum_{n=0}^{N-1}Z_n.
\end{align}
commutes with the global $\mathbb{Z}_2$ parity operator $P\coloneqq\prod_{j=0}^{N-1} Z_j$.

As another example, consider the spinful Fermi--Hubbard model
\begin{gather}
        H_{\mathrm{FH}} = -\sum_{j=0}^{N-2}\sum_{\sigma = \uparrow,\downarrow} \left(c^\dag_{j,\sigma}c_{j+1,\sigma} + c^\dag_{j+1,\sigma}c_{j,\sigma}\right) 
         + h_U \sum_{j=0}^{N-1}n_{j,\uparrow}n_{j,\downarrow}.
\end{gather}
A natural parity symmetry here is fermion-number parity,
\begin{equation}
    P \coloneqq (-1)^{\hat N}=\prod_{j,\sigma} e^{\rmi\pi n_{j,\sigma}},
    \qquad \hat N\coloneqq\sum_{j,\sigma} n_{j,\sigma}.
\end{equation}
Under the Jordan--Wigner transformation, this maps to the Pauli-$Z$ string over all spin-orbitals (qubits),
\begin{equation}
    P = \prod_{q=0}^{2N-1} Z_q .
\end{equation}

Throughout the remainder of the paper, we choose $\{|n\rangle\}_{n=0}^{d-1}$ to be a common eigenbasis of $H$ and $P$, so that
\begin{equation}
    H|n\rangle = E_n |n\rangle,\quad P|n\rangle = p_n |n\rangle,\quad p_n\in\{\pm 1\}.
\end{equation}
We say that $|n\rangle$ has even parity if $p_n=+1$ and odd parity if $p_n=-1$. Let $\mathcal{H}_{\mathrm{even}}$ and $\mathcal{H}_{\mathrm{odd}}$ denote the corresponding parity sectors (subspaces). We also define the parity projectors
\begin{equation}
    \Pi_{\rm S}\coloneqq\frac{I+P}{2},\quad \Pi_{\rm A}\coloneqq\frac{I-P}{2},
\end{equation}
which form a two-outcome parity POVM. Here $I$ denotes the identity operator.

We will use the following selection rule (see \aref{sec:proof2} for a proof).
\begin{lemma}\label{lem:parity}
    Suppose $[H,P]=0$ and $\{A,P\}=0$, and let $|m\rangle,|n\rangle$ be eigenstates of $H$ (and hence of $P$). If $\langle n|A|m\rangle\neq 0$, then $|m\rangle$ and $|n\rangle$ have opposite parity, i.e., $p_m p_n=-1$.
\end{lemma}

\subsection{Quench spectroscopy}
Quench spectroscopy (QS) is a simple yet powerful technique for probing dynamical properties of quantum systems \cite{menu18quench,despres2019correlation}. The central idea is to apply a sudden perturbation to a prepared state and then monitor the ensuing unitary dynamics under a time-independent Hamiltonian. By measuring suitable observables as a function of time, one obtains a signal whose oscillation frequencies correspond to energy differences in the spectrum. Classical post-processing of this time-domain signal then yields spectral information and dynamical correlation functions, without requiring controlled time evolution or other time-dependent operations.

Building on the QS method in Refs.~\cite{villa2019unraveling,villa2021out}, one can directly access the \emph{imaginary part} of a two-point time correlator under minimal assumptions: the initial state may be either equilibrium or nonequilibrium. To formalize the protocol, we define the \emph{quench function} as the expectation value of a measurement outcome,
\begin{equation}
Q_{\rho,M}(U) \coloneqq \mathrm{tr}(U\rho U^\dagger M),
\end{equation}
where $U$ denotes the unitary implemented in the QS experiment and $M$ is the measured observable. Assuming $A$ is a Hermitian operator satisfying $A^2 = I$, and applying the quench operator
\begin{equation}\label{eq:tqs-im}
    U_{\mathrm{Im}} \coloneqq e^{\rmi \pi A/4}=\frac{I+\rmi A}{\sqrt{2}},
\end{equation}
one obtains
    \begin{align}
          \Im[C(A,B,t)] &= -\frac{\rmi}{2}\mathrm{tr}(\rho [A,B(t)]) \nonumber \\
          &= Q_{\rho,B}\left(e^{-\rmi Ht}U_{\mathrm{Im}}\right) - \frac{1}{2}Q_{\rho,B}\left(e^{-\rmi Ht}A\right) - \frac{1}{2}Q_{\rho,B}\left(e^{-\rmi Ht}\right)\label{eq:quench_Im}.
    \end{align}
All terms on the right-hand side can be measured directly on a quantum device, making QS a practical route to extracting imaginary time-correlator information, equivalently bosonic response functions of the form $\rmi C(A,B,t)-\rmi C(B,A,t)$. The nontrivial step is to access the \emph{real part} of time correlators, or equivalently fermionic response functions of the form $C(A,B,t) + C(B,A,t)$. In the following sections, we show how to achieve this by introducing a \emph{tailored quench operator} and exploiting symmetries inherent to the system.

\section{Symmetry-enabled protocols for time correlators}
\label{sec:mainprotocol}

From now on, we focus on systems with a $\mathbb{Z}_2$ (parity) symmetry. 
Concretely, we assume there exists a parity operator $P$ such that
\begin{equation}\label{eq:assumption-p}
[P,H]=0,\quad P^\dagger=P,\quad P^2=I.
\end{equation}
We consider Hermitian observables $A$ and $B$ satisfying
\begin{equation}\label{eq:assumption-ab}
\quad A^2=I,\quad\{A,P\}=\{B,P\}=0,
\end{equation}
and we assume the initial state is parity-symmetric,
\begin{equation}\label{eq:assumption-rho}
[\rho,P]=0.
\end{equation}
Prominent examples include equilibrium (Gibbs) states $\rho_\beta=e^{-\beta H}/\mathrm{tr}(e^{-\beta H})$, and the ground state recovered in the limit $\beta\to\infty$.

\subsection{Tailored quench spectroscopy}

To access the real part of a two-point time correlator, we introduce the quench operator
\begin{equation}\label{eq:tqs-re}
    U_{\mathrm{Re}}\coloneqq\frac{P+A}{\sqrt{2}}.
\end{equation}
Since $\{P,A\}=0$ and $P^2=A^2=I$, the operator $PA$ satisfies $(PA)^\dagger=-PA$
and $(PA)^2=-I$, so $G\coloneqq-\rmi PA$ is Hermitian and obeys $G^2=I$.
Consequently, $U_{\mathrm{Re}}$ admits the decomposition
\begin{equation}
    \frac{P+A}{\sqrt{2}}
    =P\exp\!\left(\rmi\frac{\pi}{4}\,G\right)
    =P\exp\!\left[\rmi\frac{\pi}{4}\,(-\rmi PA)\right].
\end{equation}
In particular, when $P$ and $A$ are Pauli strings,
$G$ is also a Pauli string and the implementation reduces to a single Pauli rotation
(with Bloch-sphere rotation angle $\pi/2$).

We refer to the protocol based on the two quench operators $U_{\mathrm{Im}}$ and $U_{\mathrm{Re}}$ as \emph{tailored quench spectroscopy} (TQS). For an $N$-qubit system, both operators can be implemented with circuit depth $\mathcal{O}(\log N)$ using only CNOTs and single-qubit gates \cite{haah2025efficient}.

With the quench operator $U_{\mathrm{Re}}$, one obtains
    \begin{align}
        \Re[C(PA,B,t)] &= \frac{1}{2}\mathrm{tr}(\rho P \{A,B(t)\})\nonumber\\
        &= Q_{\rho,B}\left(e^{-\rmi Ht}U_{\mathrm{Re}}\right)  - \frac{1}{2}Q_{\rho,B}\left(e^{-\rmi Ht}P\right) - \frac{1}{2}Q_{\rho,B}\left(e^{-\rmi Ht}A\right)\label{eq:quench_Re}.
    \end{align}
Importantly, if the initial state $\rho$ has a well-defined parity, i.e., $P\rho = p\rho$ with $p\in\{+1,-1\}$, then $C(PA,B,t)=p\,C(A,B,t)$. Consequently, the left-hand side of \eref{eq:quench_Re} directly yields $\Re[C(A,B,t)]$ up to the known sign $p$, which can be fixed from the parity sector of the initial state. Together with the QS identity for $\Im[C(A,B,t)]$, the full complex correlator can therefore be reconstructed from a linear combination of six experimentally accessible quench functions.

Moreover, when $\{B,P\}=0$, several terms in \eref{eq:quench_Im} and \eref{eq:quench_Re} vanish identically by symmetry. For instance,
\begin{equation}
    \{P,B\} = 0 \Longrightarrow PBP = -B \Longrightarrow PB(t)P = -B(t) \Longrightarrow \mathrm{tr}(\rho B(t)) = 0.
\end{equation}
As a result, the full complex correlator can be reconstructed using only two quench functions. This simplification is summarized in the following theorem; see \aref{sec:proof3} for the proof.

\begin{theorem}\label{thm:eigenstate}
Under the assumptions \erefrange{eq:assumption-p}{eq:assumption-rho}, for all $t\in\mathbb{R}$,
\begin{equation}
\Im[C(A,B,t)] \;=\; Q_{\rho,B}\!\left(e^{-\rmi Ht}U_{\mathrm{Im}}\right).
\end{equation}
Moreover, if the initial state has definite parity, i.e., $P\rho=p\rho$ with $p\in\{\pm1\}$, then
\begin{equation}
\Re[C(A,B,t)] \;=\; p\,Q_{\rho,B}\!\left(e^{-\rmi Ht}U_{\mathrm{Re}}\right).
\end{equation}
\end{theorem}

The full protocol is demonstrated in Algorithm~\ref{alg:quench} in the form of pseudocode.

\begin{algorithm}[!htb]
        \caption{\label{alg:quench}Tailored quench spectroscopy}
        {\small
            \begin{algorithmic}[1]
            \Statex \textbf{Input:} Initial state $\rho$ with definite parity, times $t$, Hamiltonian $H$, observables $A,B$, parity operator $P$, error bound $\varepsilon$.
            \Statex \textbf{Output:} Estimate of $C(A,B,t)$ with error bound $\varepsilon$.
            \State Measure $\rho$ in the eigenbasis of $P$ to obtain its parity $\hat{p}$.
            \For{$n=1,2,\ldots, N = \caO(\varepsilon^{-2})$}
            \State Prepare $\rho$;
            \State Apply quench operator $U_{\mathrm{Im}}$ on the state $\rho$;
            \State Let the state evolve under $\exp(-\rmi Ht)$;
            \State Measure the evolved state in the eigenbasis of $B$ to obtain outcome $c^{(i)}_n$;
            \EndFor
            \State Compute $\overline{c^{(i)}} = \sum_{n=1}^{N}c^{(i)}_n/N$.
            \For{$n=1,2,\ldots, N = \caO(\varepsilon^{-2})$}
            \State Prepare $\rho$;
            \State Apply quench operator $U_{\mathrm{Re}}$ on $\rho$;
            \State Let the state evolve under $\exp(-\rmi Ht)$;
            \State Measure the evolved state in the eigenbasis of $B$ to obtain outcome $c^{(r)}_n$;
            \EndFor
            \State Compute $\overline{c^{(r)}} = \sum_{n=1}^N c^{(r)}_n/N$.
            \State Return $\hat{p}\,\overline{c^{(r)}} + \rmi\,\overline{c^{(i)}}$.
            \end{algorithmic}
        }
    \end{algorithm}

Theorem~\ref{thm:eigenstate} applies in particular when the initial state $\rho$ is a non-degenerate eigenstate of $H$, in which case $\rho$ has a definite parity $p\in\{\pm 1\}$. In the degenerate case, since $[H,P]=0$, one can choose an energy eigenbasis that simultaneously diagonalizes $P$ within each degenerate eigenspace, and hence prepare a \emph{pure} eigenstate with definite parity. In practice, parity can be assessed by estimating $\langle P\rangle_\rho=\mathrm{tr}(\rho P)$ using standard end-of-circuit measurements. If parity leakage is observed (i.e., $|\langle P\rangle_\rho|<1$), one may, for example, enforce a parity-preserving state-preparation circuit so that the prepared state remains within the desired parity sector. Since $[H,P]=0$, parity is conserved under time evolution, so this calibration can be performed once and reused for subsequent quench spectroscopy runs.

Next, we show that an analogous strategy applies to thermal states. Consider a finite-temperature thermal (quantum Gibbs) state
\begin{equation}
    \rho_\beta \coloneqq \frac{e^{-\beta H}}{\mathcal{Z}},
\qquad 
\mathcal{Z} \coloneqq \mathrm{tr}(e^{-\beta H}),
\end{equation}
at inverse temperature $\beta$. Since $[H,P]=0$, it follows that $[\rho_\beta,H]=[P,\rho_\beta]=0$. However, $\rho_\beta$ generally does not have definite parity, i.e., it does not satisfy the stronger condition
\begin{equation}
P\rho_\beta = p\,\rho_\beta,\qquad p\in\{\pm1\}.
\end{equation}
To recover a fixed-parity setting, we introduce the \emph{symmetric and antisymmetric thermal states}, which correspond to the thermal state projected onto the even and odd parity sectors:
\begin{equation}\label{eq:sym_thermal_state}
    \begin{split}
            \rho_\mathrm{S} \coloneqq \frac{\Pi_\mathrm{S}\rho_\beta\Pi_\mathrm{S}}{\mathrm{tr}(\rho_\beta\Pi_\mathrm{S})}
            = \frac{\rho_\beta + \rho_\beta P}{1 + \mathrm{tr}(\rho_\beta P)},\quad
            \rho_\mathrm{A} \coloneqq \frac{\Pi_\mathrm{A}\rho_\beta\Pi_\mathrm{A}}{\mathrm{tr}(\rho_\beta\Pi_\mathrm{A})} = \frac{\rho_\beta - \rho_\beta P}{1 - \mathrm{tr}(\rho_\beta P)},
    \end{split}
\end{equation}
where $\Pi_\mathrm{S}=(I+P)/2$ and $\Pi_\mathrm{A}=(I-P)/2$ are two projectors: $\Pi_\rmS^2 = \Pi_\rmS, \Pi_\rmA^2 = \Pi_\rmA$. Both $\rho_\rmS,\rho_\rmA$ are valid quantum states with trace 1.

These satisfy $P\rho_\mathrm{S}=\rho_\mathrm{S}$ and $P\rho_\mathrm{A}=-\rho_\mathrm{A}$, and the original thermal state decomposes as
\begin{equation}
    \rho_\beta
    =\frac{1 + \mathrm{tr}(\rho_\beta P)}{2}\rho_\mathrm{S}
    +\frac{1 - \mathrm{tr}(\rho_\beta P)}{2}\rho_\mathrm{A}.
\end{equation}

With these definitions, TQS provides an efficient route to estimating the real component of $C(A,B,t)$, as summarized in the following theorem; see \aref{app:proofofthm2} for its proof.
\begin{theorem}\label{thm:thermalstate}
Let $\rho_\beta$, $\rho_\mathrm{S}$, and $\rho_\mathrm{A}$ denote the thermal, symmetric, and antisymmetric thermal states of $H$ at inverse temperature $\beta$. Under the assumptions \eqsref{eq:assumption-p}{eq:assumption-ab}, for all $t\in\mathbb{R}$,
\begin{align}
    \Re[C(A,B,t)]
    &=
    \frac{1 + \mathrm{tr}(\rho_\beta P)}{2}\,
    Q_{\rho_\mathrm{S},B}\!\left(e^{-\rmi Ht}U_{\mathrm{Re}}\right)
    \nonumber\\
    &
    -\frac{1 - \mathrm{tr}(\rho_\beta P)}{2}\,
    Q_{\rho_\mathrm{A},B}\!\left(e^{-\rmi Ht}U_{\mathrm{Re}}\right),
    \label{eq: theorem2}
\end{align}
and the imaginary component can be estimated analogously to Theorem~\ref{thm:eigenstate}.
\end{theorem}

\subsection{Preparation of \texorpdfstring{$\rho_\mathrm{S},\rho_\mathrm{A}$}{rho S and rho A}}

Since thermal-state preparation itself is a challenging task on near-term hardware, we treat the generation of parity-resolved Gibbs states as an input primitive: our TQS protocol applies to any procedure that can prepare $\rho_{\mathrm S}$ and $\rho_{\mathrm A}$, exactly or approximately. Nevertheless, we outline two possible approaches. Measuring $\rho_\beta$ with the two-outcome POVM $\{\Pi_{\mathrm S},\Pi_{\mathrm A}\}$ provides a conceptually simple way to obtain $\rho_{\mathrm S}$ and $\rho_{\mathrm A}$. However, since $P$ and $B$ anticommute, a direct parity measurement is not compatible with the subsequent measurement in the eigenbasis of $B$ within the same experimental run without additional overhead (e.g., ancilla-assisted parity checks with feed-forward and/or post-selection). Moreover, mid-circuit measurements are typically slow and noisy, making this approach less attractive. 

Alternatively, we introduce a tailored \emph{parity-preserving quantum sampler} based on dissipative state preparation, which can directly prepare parity-resolved Gibbs states without requiring mid-circuit measurements. We first briefly review a Lindbladian framework for preparing Gibbs states via dissipative evolution that has recently attracted renewed interest
\cite{chen2025efficient,tong2025fast,ding2025efficient,Rouze2026,Ding2026}. Let $\mathcal{L}$ be a Lindbladian and consider the
master equation
\begin{equation}
\frac{d}{dt}\rho(t)=\mathcal{L}[\rho(t)].
\end{equation}
Let $\rho(t) = e^{\caL t}[\rho(0)]$ denote the solution of the master equation. If the map $e^{\caL t}$ is \textit{ergodic} \cite{breuer2002theory}, then there exists a unique steady state (in the context of quantum Gibbs sampler, the steady state is $\rho_\beta$) such that $\lim_{t\to\infty}\rho(t) = \rho_\beta$ for any initial
state $\rho(0)$. In this way, Gibbs-state preparation is reduced to simulating dissipative dynamics.

A canonical construction is the \emph{Davies generator} \cite{Lindblad1976,GoriniKossakowskiSudarshan1976}. Fix a system operator $J$ and define its
\emph{Bohr-frequency components} (i.e., components labeled by energy gaps $\nu=E_m-E_n$) by
\begin{equation}
J_\nu \coloneqq \lim_{T\to\infty}\frac{1}{2T}\int_{-T}^T dt\; e^{\rmi Ht}J e^{-\rmi Ht}\,e^{-\rmi\nu t},
\end{equation}
so that $[H,J_\nu] = \nu J_\nu,e^{\rmi Ht}Je^{-\rmi Ht}=\sum_\nu J_\nu e^{\rmi\nu t}$.
Let $\eta(\nu)$ be any function satisfying the relation $\eta(-\nu)=e^{\beta\nu}\eta(\nu)$. Then the Davies generator is defined as 
\begin{equation}\label{eq:davies}
\mathcal{L}_D[\cdot]
=\sum_\nu \eta(\nu)\!\left[J_\nu(\cdot)J_\nu^\dagger-\frac{1}{2}\{J_\nu^\dagger J_\nu,\cdot\}\right].
\end{equation}
This Davies generator satisfies the Kubo-Martin-Schwinger (KMS) detailed balance condition \cite{ding2025efficient} with respect to $\rho_\beta$:
\begin{equation}
    \caL_D[\cdot] = \rho_\beta^{1/2}\caL^\dag_D\left[\rho_\beta^{-1/2}\cdot \rho_\beta^{-1/2}\right]\rho_\beta^{1/2}.
\end{equation}
This implies $\mathcal{L}_D[\rho_\beta]=0$. A self-consistent derivation is included in \aref{app:davies}.

To ensure that $\rho_\beta$ is the \emph{unique} steady state on the full Hilbert space, one imposes an irreducibility condition. One convenient formulation is via reducing projectors: a projector $\Pi$ \emph{reduces} $\mathcal{L}_D$ if the corresponding subspace is invariant under the
dynamics, namely
\begin{equation}\label{eq:reducible}
\mathcal{L}_D[\Pi\rho\Pi]=\Pi\,\mathcal{L}_D[\Pi\rho\Pi]\,\Pi,\quad \forall \rho.
\end{equation}
If the only reducing projectors are $\Pi=0$ and $\Pi=I$, then $\mathcal{L}_D$ is \emph{irreducible}
\cite{zhang2024criteria}; under additional mild assumptions (e.g., full rank of the stationary state), this implies
a unique full-rank stationary state, which must be $\rho_\beta$ by detailed balance.

In our application, however, we do not need global irreducibility. Instead, we want the dynamics to
\emph{preserve parity} so that it does not mix the even and odd sectors, and then be ergodic within each
sector. This yields the parity-resolved Gibbs states $\rho_{\rm S}$ and $\rho_{\rm A}$ as the long-time limits.

Assuming $[H,P]=0$, if we choose a jump operator that commutes with parity, $[J,P]=0$, then the Davies dynamics is block-diagonal in the parity decomposition.

\begin{proposition}\label{prop:parity-reducing}
Suppose $[H,P]=0$ and $[J,P]=0$. Then both $\Pi_{\rm S}$ and $\Pi_{\rm A}$ reduce the Davies generator
$\mathcal{L}_D$.
\end{proposition}

\begin{proof}
Since $[H,P]=0$, we have $e^{\rmi Ht}Pe^{-\rmi Ht}=P$. If $[J,P]=0$, then
$[J(t),P]=0$ for all $t$, where $J(t)=e^{\rmi Ht}Je^{-\rmi Ht}$. Taking Fourier components implies
$[J_\nu,P]=0$ for all $\nu$, hence $[J_\nu,\Pi_{\rm S}]=[J_\nu,\Pi_{\rm A}]=0$.

Let $\sigma$ be supported in the even sector, i.e., $\sigma=\Pi_{\rm S}\sigma\Pi_{\rm S}$. Then for each $\nu$,
\begin{align}
J_\nu \sigma J_\nu^\dagger
&=J_\nu(\Pi_{\rm S}\sigma\Pi_{\rm S})J_\nu^\dagger
=\Pi_{\rm S}(J_\nu\sigma J_\nu^\dagger)\Pi_{\rm S},\\
J_\nu^\dagger J_\nu \sigma
&=J_\nu^\dagger J_\nu(\Pi_{\rm S}\sigma\Pi_{\rm S})
=\Pi_{\rm S}(J_\nu^\dagger J_\nu\sigma)\Pi_{\rm S},
\end{align}
and similarly for $\sigma J_\nu^\dagger J_\nu$. Plugging these relations into \eref{eq:davies} yields
$\mathcal{L}_D[\sigma]=\Pi_{\rm S}\mathcal{L}_D[\sigma]\Pi_{\rm S}$, i.e., $\Pi_{\rm S}$ reduces $\mathcal{L}_D$.
The same argument applies to $\Pi_{\rm A}$.
\end{proof}
The proposition implies that there is no transition between $\caH_{\mathrm{even}}$ and $\caH_{\mathrm{odd}}$ in the dissipative dynamics.
Because the parity sectors are invariant, the long-time limit of the dynamics depends on the initial parity
weights. In particular, if the restriction of $\mathcal{L}_D$ to $\mathcal{H}_{\rm even}$ is ergodic, then the
unique stationary state within that sector is the even-parity Gibbs state $\rho_{\rm S}$; likewise, ergodicity on
$\mathcal{H}_{\rm odd}$ yields $\rho_{\rm A}$. Thus, by preparing an initial state supported in the desired sector
and evolving under a parity-preserving Davies generator that is ergodic within that sector, we obtain
$\rho_{\rm S}$ or $\rho_{\rm A}$ without mid-circuit parity measurements. Here we emphasize that the ergodicity condition within the parity sectors is not proved in our parity-preserving Gibbs sampler.

In practice, Lindbladian evolution of this type can be implemented using dissipative dynamics and related simulation primitives \cite{ding2024single,ding24simulating,yu2025lindbladian}. While some implementations introduce additional ancilla qubits, they avoid long-range controlled unitaries and deep interference structures, making them potentially suitable for near-term or early fault-tolerance quantum devices. 

We now illustrate how to construct a parity-preserving Lindbladian whose unique steady state is the symmetric thermal state. As a concrete example, we consider the $N=4$ site XXZ model in \eref{eq:XXZ} with parameters $J_X=2$, $J_Z=1$, and $h=10$. To generate the full thermal state $\rho_\beta$, we choose a parity-breaking jump operator $J=X_0$. To generate the symmetric thermal state $\rho_\mathrm{S}$, we instead choose the parity-preserving two-site operator $J=Y_0Y_1$. We denote the corresponding Davies generators by $\mathcal{L}_D$ (for $J=X_0$) and $\mathcal{L}_{D,\mathrm{S}}$ (for $J=Y_0Y_1$).

\begin{figure}[!htb]
    \centering
    \includegraphics[width=0.6\linewidth]{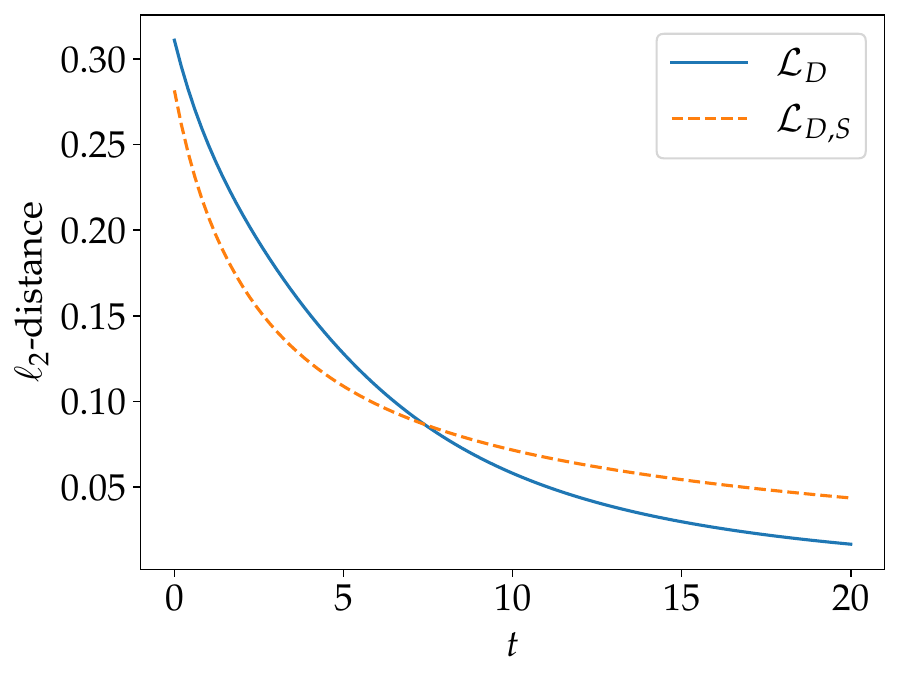}
    \caption{Preparation of thermal states using Davies dynamics for the finite-size XXZ example. The horizontal axis is the Lindbladian evolution time \(t\), and the vertical axis is the \(\ell_2\)-distance between the vectorized evolved state and the corresponding steady state: $\|v - w\|_2 \coloneqq (\sum_i |v_i - w_i|^2)^{1/2}$, which equals the Frobenius distance between density operators divided by $\sqrt{d}$. The blue solid curve corresponds to \(\mathcal L_D\) and convergence to the full Gibbs state, whereas the orange dashed curve corresponds to the parity-preserving generator \(\mathcal L_{D,\mathrm S}\) and convergence to the even-parity Gibbs state \(\rho_{\mathrm S}\). The initial state is a random state supported in the even-parity sector. The relative convergence rates and the crossing of the curves are specific to this choice of generators and initial state and are not intended to imply a generic mixing-time advantage or disadvantage of the parity-preserving construction.}
    \label{fig:sym_davies}
\end{figure}

To simulate the Lindbladian dynamics, we vectorize the density matrix using the Pauli basis. Let $\mathcal{P}_n=\{P_i\}_{i=0}^{d^2-1}$ be the set of $n$-qubit Pauli operators (including the identity), where $d=2^n$. We define the vectorization of $\rho$ as
\begin{equation}
    |\rho\>\!\> = \frac{1}{d}\sum_{i=0}^{d^2-1}\mathrm{tr}(\rho P_i)\boldsymbol{e}_i,
\end{equation}
with $\{\boldsymbol{e}_i\}_{i=0}^{d^2-1}$ the standard basis of $\bbR^{d^2}$. Under this mapping, the action of a Lindbladian $\mathcal{L}$ becomes a matrix $\hat{\mathrm{L}}$ acting linearly on $|\rho\>\!\>$, i.e.,
\begin{equation}
\begin{aligned}
|\caL(\rho)\>\!\> &= \frac{1}{d}\sum_{i=0}^{d^2-1}\mathrm{tr}(\caL(\rho) P_i)\boldsymbol{e}_i
= \frac{1}{d}\sum_{i=0}^{d^2-1}\mathrm{tr}(\rho \caL^\dag(P_i))\boldsymbol{e}_i = \hat{\mathrm{L}}|\rho\>\!\>,
\end{aligned}
\end{equation}
where
\begin{align}
    \hat{\mathrm{L}} &= \frac{1}{d}\sum_{i,j=0}^{d^2-1}\mathrm{tr}(\caL^\dag(P_i)P_j)\boldsymbol{e}_i \boldsymbol{e}_j^\top = \frac{1}{d}\sum_{i,j=0}^{d^2-1}\mathrm{tr}(P_i\caL(P_j))\boldsymbol{e}_i \boldsymbol{e}_j^\top.
\end{align}
The Lindbladian equation thus becomes the linear ODE
\begin{equation}
    \frac{d}{dt}|\rho(t)\>\!\> = \hat{\mathrm{L}}|\rho(t)\>\!\>.
\end{equation}
We initialize \(\rho(0)\) as a random state supported in the even-parity subspace and evolve it under the two Davies generators. As shown in Fig.~\ref{fig:sym_davies}, both dynamics converge to their corresponding stationary states, with \(\mathcal L_{D,\mathrm S}\) converging to the parity-resolved thermal state \(\rho_{\mathrm S}\).

We emphasize that the relative convergence rates in Fig.~\ref{fig:sym_davies} are specific to this finite-size example and should not be interpreted as a generic property of the parity-preserving construction. Parity preservation ensures invariance of the symmetry sectors, but does not by itself imply faster or slower mixing. The convergence behavior is instead determined by the spectral properties of the corresponding Lindbladian and depends on the choice of jump operators and transition rates.

The crossing of the two curves likewise has no universal significance. In general, the transient convergence curve depends on the overlap of the initial state with the decay modes of the Lindbladian. Since the initial state used here is a particular random state in the even-parity sector, changing this state can modify the transient behavior and hence the location, or even the presence, of such a crossing. The purpose of Fig.~\ref{fig:sym_davies} is therefore to demonstrate the feasibility of preparing a parity-resolved Gibbs state using parity-preserving Davies dynamics, rather than to establish an improvement in mixing time.

\subsection{Generalization to OTOC}
\label{sec:otoc}

Our approach can be partially extended to a class of $n$-point correlators \cite{peder2014efficient,del2024robust,wang2025computing}. Let $O_0,O_1,\ldots,O_{n-1}$ be $n$ observables and consider the correlator
\begin{equation}
    \mathrm{tr}(\rho\, O_0 O_1 \cdots O_{n-1}).
\end{equation}
Assume the initial state is parity-symmetric, $[\rho,P]=0$, and that $O_0$ is parity-odd, $\{O_0,P\}=0$. As in the two-point setting, we introduce the tailored quench operators
\begin{equation}
U_{\mathrm{Im}}=\frac{I+\rmi O_0}{\sqrt{2}},\qquad
U_{\mathrm{Re}}=\frac{P+O_0}{\sqrt{2}}.
\end{equation}

We focus on the case where the remaining operators $O_1,\ldots,O_L$ are Heisenberg evolutions of Pauli observables, i.e., $O_\ell = O_\ell(t_\ell)=e^{\rmi Ht_\ell} O_\ell e^{-\rmi Ht_\ell}$, and the ordered product is invariant under reversal,
\begin{equation}
    O_1 O_2 \cdots O_{n-1} = O_{n-1} \cdots O_2 O_1,
\end{equation}
which holds, for example, for certain palindromic strings arising in out-of-time-order correlators. Under these conditions, the same TQS measurement primitives can be used to access the desired correlator.

\begin{figure*}[!htb]
    \centering
    \includegraphics[width=0.9\linewidth]{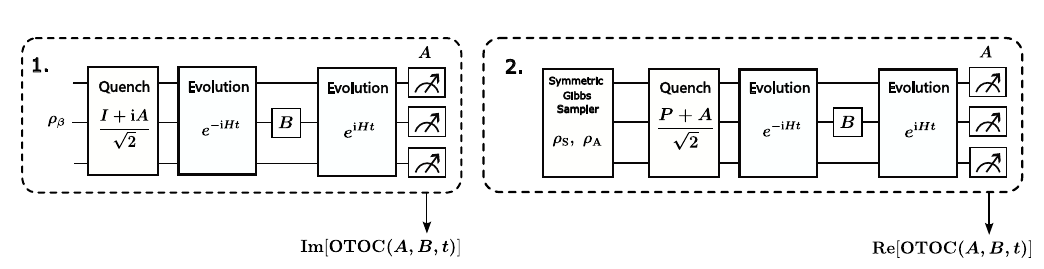}
    \caption{Quantum circuits for estimating $\otoc(A,B,t)$ using tailored quench spectroscopy.}
    \label{fig:otoc}
\end{figure*}

A particularly important example is the out-of-time-order correlator (OTOC), widely used to diagnose quantum chaos and information scrambling \cite{swingle16measuring,xu24scrambling,hashimoto2017out,garcia2022out}. For a thermal state $\rho_{\beta}$ and Pauli operators $A,B$, we define
\begin{equation}
        \otoc(A,B,t) = \mathrm{tr}\!\big[\rho_{\beta}\,A\,B(t)\,A\,B(t)\big],
\end{equation}
where $B(t) =e^{\rmi Ht}Be^{-\rmi Ht}$. 
Introduce the operator $\widetilde{A}\coloneqq B e^{-\rmi Ht} A e^{\rmi Ht}B$. Then its Heisenberg evolution satisfies
\begin{equation}
\widetilde{A}(t)=e^{\rmi Ht}\widetilde{A}e^{-\rmi Ht}=B(t)\,A\,B(t),
\end{equation}
and therefore
\begin{equation}
\otoc(A,B,t)=\mathrm{tr}\!\big[\rho_\beta\,A\,\widetilde{A}(t)\big]\coloneqq C(A,\widetilde{A},t),
\end{equation}
where we use the convention $C(X,Y,t)=\mathrm{tr}(\rho_\beta\,X\,Y(t))$. Experimentally, this reduces to estimating the modified quench functions
\begin{gather}
    Q_{\rho_\beta,A}\!\left(e^{\rmi Ht} B e^{-\rmi Ht}U_{\mathrm{Im}}\right),\quad Q_{\rho_\rmA,A}\!\left(e^{\rmi Ht} B e^{-\rmi Ht}U_{\mathrm{Re}}\right),\quad Q_{\rho_\rmS,A}\!\left(e^{\rmi Ht} B e^{-\rmi Ht}U_{\mathrm{Re}}\right).
\end{gather}
The full quantum circuit is shown in \fref{fig:otoc}.

\section{Complexity and error analysis}
\label{sec:analysis}

\subsection{Complexity analysis}

For a fixed time $t$, estimating $C(A,B,t)$ (or its real/imaginary parts) reduces to estimating a constant number of quench functions. Therefore, to achieve additive error $\varepsilon$ with constant success probability, the number of shots per time point scales as $\mathcal{O}(\varepsilon^{-2})$ by standard concentration bounds.

To reconstruct the Green's function, we first
estimate the active transition frequencies
from the measured time-domain correlator. Suppose the correlator contains $\chi$
distinct transition frequencies
$\{\Omega_k\}_{k=1}^{\chi}$ with nonzero
spectral weights. Define their minimum separation
by
\begin{equation}
    \Delta_{\rm spec}
    =
    \min_{k\ne l}|\Omega_k-\Omega_l|.
\end{equation}
For uniform sampling with interval $\delta t$,
the dimensionless frequencies are
$f_k=\Omega_k\delta t\pmod{2\pi}$.
The sampling interval must be chosen to avoid
aliasing of the relevant frequency range.

In the standard separation-limited regime,
a sufficiently long observation window scales as
\begin{equation}
    T_{\max}
    =O(\Delta_{\rm spec}^{-1}),
\end{equation}
under suitable separation and noise assumptions.
The corresponding number of sampling times is
\begin{equation}
    M=O\left(
    \frac{1}{\Delta_{\rm spec}\delta t}
    \right).
\end{equation}
Estimating each correlator to additive accuracy
$\varepsilon$ with bounded measurement outcomes
requires
$O(\varepsilon^{-2}\log(M/\delta))$
circuit executions per time point to achieve
simultaneous accuracy with probability at least
$1-\delta$. Consequently, the total number
of circuit executions is
\begin{equation}
    O\left(
    \frac{\varepsilon^{-2}}
    {\Delta_{\rm spec}\delta t}
    \log\frac{M}{\delta}
    \right).
\end{equation}
This is a correlator-sampling cost; guaranteeing
a specified frequency-estimation accuracy
requires additional signal-conditioning
assumptions.

For thermal correlators, the active frequencies
are Bohr frequencies $E_m-E_n$ with nonzero
spectral weights. Their minimum separation is
generally distinct from the smallest nonzero
energy-level gap.

\subsection{Error analysis}

In the nonequilibrium setting, an error in the prepared initial state translates directly into an error in the correlator: since $C(A,B,t)$ is linear in $\rho$, the deviation is bounded by the trace distance $\|\rho-\rho_{\mathrm{ideal}}\|_1$ up to an operator-norm factor. In the equilibrium setting, we can further suppress symmetry-breaking preparation errors by applying a \emph{symmetrization channel}
\begin{equation}
\mathcal{S}(\rho)=\frac{\rho+P\rho P}{2},
\end{equation}
which removes coherences between opposite parity sectors.

To illustrate the effect, suppose the ideal initial state is the ground state $|0\rangle$ and the actual prepared state is
\begin{equation}
|\psi\rangle=\sqrt{1-|\mu|^2}|0\rangle+\mu|1\rangle,
\qquad |\mu|\ll 1,
\end{equation}
where $|1\rangle$ has opposite parity to $|0\rangle$. Consider the imaginary component $\Im[C(A,B,t)]$ with inaccurate input state $|\psi\>$, which can be expanded as
\begin{align}
Q_{|\psi\>,B}(e^{-\rmi Ht} e^{\rmi \pi A/4}) =
\frac{1}{2}\langle\psi|B(t)|\psi\rangle
+\frac{1}{2}\langle\psi|A B(t)A|\psi\rangle 
-\frac{1}{2}\langle\psi|\, \rmi[A,B(t)]\,|\psi\rangle.
\end{align}
Note that $\rmi[A,B(t)]$ is parity-even (symmetry-preserving), whereas $B(t)$ and $AB(t)A$ are parity-odd (symmetry-breaking) under the assumptions $\{A,P\}=\{B,P\}=0$.  Hence, the error between the ideal case and the actual case is
\begin{align}\label{eq:err}
&\quad Q_{|\psi\>,B}(e^{-\rmi Ht} e^{\rmi \pi A/4}) - Q_{|0\>,B}(e^{-\rmi Ht} e^{\rmi \pi A/4}) \nonumber\\
&= \sqrt{1 - |\mu|^2}(\Re\!\big[\mu^*\langle 1|B(t)|0\rangle\big]
+\Re\!\big[\mu^*\langle 1|AB(t)A|0\rangle\big]) \nonumber\\
&\quad -\frac{|\mu|^2}{2}\langle 1|\, \rmi[A,B(t)]\,|1\rangle
+\frac{|\mu|^2}{2}\langle 0|\, \rmi[A,B(t)]\,|0\rangle .
\end{align}
The leading contribution is $\mathcal{O}(|\mu|)$.

After applying symmetrization, the initial state becomes the incoherent mixture
\begin{equation}
\mathcal{S}(|\psi\rangle\langle\psi|)
= (1-|\mu|^2)|0\rangle\langle 0|+|\mu|^2|1\rangle\langle 1|,
\end{equation}
which removes the off-diagonal coherence terms responsible for the $\mathcal{O}(|\mu|)$ contribution. As a result, the leading error is reduced to $\mathcal{O}(|\mu|^2)$. The analysis for the real component $\Re[C(A,B,t)]$ is similar.

\section{Numerical simulation}
\label{sec:numeric}
In this section, we present numerical experiments validating the TQS protocol and its robustness to state-preparation errors. To simulate real-time dynamics without ancillas, we use a first-order Trotter--Suzuki (TS) product formula \cite{childs2019nearly,childs2021theory,tran21faster}. For a time-independent Hamiltonian $H=\sum_{\gamma=1}^{\Gamma} H_\gamma$, where each $H_\gamma$ is a sum of mutually commuting Pauli terms, we approximate the time-evolution operator by
\begin{equation}
U_{\mathrm{TS}}(t)
\coloneqq\left[\prod_{\gamma=1}^{\Gamma}\exp\!\left(-\rmi H_\gamma\, t/N_T\right)\right]^{N_T},
\end{equation}
where $N_T$ is the number of Trotter steps (circuit layers) and directly controls circuit depth. For fixed $t$, the first-order TS error typically decreases as $\mathcal{O}(\norm{H}t^2/N_T)$. Higher-order product formulas can further improve the asymptotic scaling in $t$ and $N_T$ \cite{childs2019nearly,childs2021theory}. In our setting, if each term $H_\gamma$ respects parity, i.e., $[H_\gamma,P]=0$, then $U_{\mathrm{TS}}(t)$ also commutes with $P$; consequently, the simulated dynamics preserves parity and the symmetry-based conclusions of the previous sections continue to apply. 

Because the initial state is the ground state, both the eigenstate condition and the low-energy state condition help improve the error scaling \cite{Sahinoglu2021LowEnergy,Yi2022Spectral}. Let $U_{\rm TS}(t)$ denote the TS simulation operator with Trotter steps $N_T$. Define
\begin{align}
C_{\mathrm{TS}}(A,B,t) \coloneqq
\<0|A U_{\rm TS}^\dag (t)B U_{\rm TS}(t)|0\>.
\end{align}
Therefore,
\begin{equation}
\label{eq:state_dependent_corr_bound}
\begin{aligned}
    |C_{\mathrm{TS}}(A,B,t)-C(A,B,t)| &\le |\<0| A U^\dag(t)B U(t)|0\> - \<0| A U_{\rm TS}^\dag(t)BU_{\rm TS}(t)|0\>| \\
    &\le |\<0| A U^\dag(t)BU(t)|0\> - \<0| A U^\dag(t) BU_{\rm TS}(t)|0\>|  \\
    &\quad  + |\<0| A U^\dag (t)BU_{\rm TS}(t)|0\> - \<0| A U^\dag_{\rm TS}(t)BU_{\rm TS}(t)|0\>| \\
 & \le
\|B\|(\epsilon_0(t)+\epsilon_1(t)) = \epsilon_0(t) + \epsilon_1(t),
\end{aligned}
\end{equation}
where
\begin{align}
\epsilon_0(t)
\coloneqq
\left\|
\bigl(U_{\mathrm{TS}}(t)-U(t)\bigr)|0\rangle
\right\|,\quad 
\epsilon_1(t)
\coloneqq
\left\|
\bigl(U_{\mathrm{TS}}(t)-U(t)\bigr)A|0\rangle
\right\|.
\end{align}
Since $|0\>$ is an eigenstate, we can use the result of Ref. \cite{Yi2022Spectral}. In many cases including the two-layer scenario $H = H_{\gamma_1} + H_{\gamma_2}$, we have the condition $\sum_{\gamma > \gamma'}\<0|[H_{\gamma},H_{\gamma'}]|0\> = 0$, which leads to
\begin{equation}
    \epsilon_0(t) = O(t^3/N_T^2) + O(t/N_T),
\end{equation}
Compared with the worst-case error bound, there is an improvement of order $O(t/N_T)$. Furthermore, suppose $A|0\>$ is supported in a low-energy subspace, Ref.~\cite{Sahinoglu2021LowEnergy} gives a constant improvement in the error bound of $\epsilon_1(t)$.

To model imperfect initial-state preparation, we generate a random mixed-state perturbation as follows. We first draw a random traceless Hermitian matrix $E$ by sampling independent Gaussian entries (and enforcing Hermiticity), and normalize it to $\widetilde{E}$ such that $\|\widetilde{E}\|_1=1$. We then define a full-rank state
$\varrho \coloneqq e^{\widetilde{E}}/\mathrm{tr}(e^{\widetilde{E}})$.
Given a noise strength $\varepsilon\in[0,1]$, we perturb the target initial state $\rho$ by forming $\rho+\varepsilon \varrho$ and renormalizing to trace one, yielding $\widetilde{\rho}$. This construction produces a controlled trace-distance deviation, with $\|\rho-\widetilde{\rho}\|_1$ typically of order $\varepsilon$.

For the Green's function demonstrations, we consider the Fermi--Hubbard model (FHM) on a $2\times 3$ lattice with open boundary conditions, as depicted in \fref{fig:FH23}. Lattice sites are labeled by $(x,y)$ with $x\in\{0,1\}$ and $y\in\{0,1,2\}$, giving $12$ spin-orbitals, which we index according to
\begin{equation}
    (x,y,\uparrow)\mapsto 6x+2y,\qquad (x,y,\downarrow)\mapsto 6x+2y+1 .
\end{equation}
In units of the hopping amplitude, the Hamiltonian reads
\begin{equation}
    H_{\mathrm{FH}} = -\sum_{\langle p,q\rangle}\sum_{\sigma\in\{\uparrow,\downarrow\}}\left(c_{p\sigma}^\dag c_{q\sigma} + c_{q\sigma}^\dag c_{p\sigma}\right) + h_U\sum_{p} n_{p\uparrow}\,n_{p\downarrow},
\end{equation}
where $\langle p,q\rangle$ runs over nearest-neighbor pairs of sites and $h_U$ denotes the on-site interaction strength.

We work with the rescaled Hamiltonian $H \coloneqq \gamma H_{\mathrm{FH}}$ and decompose it for TS simulation as $H = \sum_{k=1}^{4} H_{\gamma_k}$, where
\begin{equation}
\begin{aligned}
    H_{\gamma_1} &= -\gamma \sum_{j\in\{0,1,6,7\}}\left(c_{j}^\dag c_{j+2} + c_{j+2}^\dag c_{j}\right),\\
    H_{\gamma_2} &= -\gamma \sum_{j\in\{2,3,8,9\}}\left(c_{j}^\dag c_{j+2} + c_{j+2}^\dag c_{j}\right),\\
    H_{\gamma_3} &= -\gamma \sum_{j=0}^{5}\left(c_{j}^\dag c_{j+6} + c_{j+6}^\dag c_{j}\right),\\
    H_{\gamma_4} &= \gamma\, h_U\sum_{j=0}^{5} n_{2j}\,n_{2j+1}.
\end{aligned}
\end{equation}
The terms within each layer $H_{\gamma_k}$ act on disjoint sets of modes (or, for $k=4$, are mutually commuting number operators). Hence each $e^{-\rmi t H_{\gamma_k}}$ factorizes exactly into a product of local unitaries.

The rescaling factor is chosen as
\begin{equation}
    \gamma \coloneqq \frac{\pi}{\Lambda},\qquad \Lambda \geq \|H_{\mathrm{FH}}\|,
\end{equation}
where $\Lambda = \|H_{\mathrm{FH}}\|$ if the spectral norm is known exactly. Otherwise $\Lambda$ is any efficiently computable upper bound, e.g., one obtained from the triangle inequality applied to the terms of $H_{\mathrm{FH}}$. This choice guarantees $\|H\|\leq\pi$, i.e., $\operatorname{spec}(H)\subseteq[-\pi,\pi]$. Consequently, each eigenphase $\theta\in\operatorname{spec}(H)$ is uniquely determined by $e^{\rmi\theta}$, up to the identification of the endpoints $\theta=\pm\pi$. This ambiguity is excluded whenever the bound is strict, $\Lambda>\|H_{\mathrm{FH}}\|$.

\begin{figure}[!htb]
\centering
\begin{tikzpicture}[
    scale=1.1,
    site/.style={
        circle,draw,fill=white,
        inner sep=2pt,
        minimum size=15pt,
        font=\scriptsize
    },
    hop1/.style={thick,blue!60},
    hop2/.style={thick,red!60,dashed},
    hop3/.style={thick,green!60!black},
    inter/.style={thick,orange!85!black,densely dotted}
]



\node[site] (u00) at (-0.34,3.24) {0};
\node[site] (u01) at (-0.34,1.74) {2};
\node[site] (u02) at (-0.34,0.24) {4};

\node[site] (u10) at (2.86,3.24) {6};
\node[site] (u11) at (2.86,1.74) {8};
\node[site] (u12) at (2.86,0.24) {10};

\node[site] (d00) at (0.34,2.76) {1};
\node[site] (d01) at (0.34,1.26) {3};
\node[site] (d02) at (0.34,-0.24) {5};

\node[site] (d10) at (3.54,2.76) {7};
\node[site] (d11) at (3.54,1.26) {9};
\node[site] (d12) at (3.54,-0.24) {11};

\draw[hop1] (u00) -- (u01);
\draw[hop1] (u10) -- (u11);
\draw[hop1] (d00) -- (d01);
\draw[hop1] (d10) -- (d11);

\draw[hop2] (u01) -- (u02);
\draw[hop2] (u11) -- (u12);
\draw[hop2] (d01) -- (d02);
\draw[hop2] (d11) -- (d12);

\draw[hop3] (u00) -- (u10);
\draw[hop3] (u01) -- (u11);
\draw[hop3] (u02) -- (u12);

\draw[hop3] (d00) -- (d10);
\draw[hop3] (d01) -- (d11);
\draw[hop3] (d02) -- (d12);


\draw[inter] (u00) -- (d00);
\draw[inter] (u01) -- (d01);
\draw[inter] (u02) -- (d02);

\draw[inter] (u10) -- (d10);
\draw[inter] (u11) -- (d11);
\draw[inter] (u12) -- (d12);

\node[font=\scriptsize] at (0,3.95) {$x=0$};
\node[font=\scriptsize] at (3.2,3.95) {$x=1$};

\node[font=\scriptsize,anchor=east] at (-0.95,3) {$y=0$};
\node[font=\scriptsize,anchor=east] at (-0.95,1.5) {$y=1$};
\node[font=\scriptsize,anchor=east] at (-0.95,0) {$y=2$};

\end{tikzpicture}

\vspace{0.5em}

\footnotesize
\noindent
$H_{\gamma_1}$: {\color{blue!60}solid blue} bonds;\quad
$H_{\gamma_2}$: {\color{red!60}dashed red} bonds;\quad
$H_{\gamma_3}$: {\color{green!60!black}solid green} spin-preserving bonds;\quad
$H_{\gamma_4}$: {\color{orange!85!black}dotted orange} on-site interactions.

\caption{
Fermi--Hubbard model on a $2\times3$ lattice.
Each physical site $(x,y)$ contains two spin orbitals,
mapped to fermionic mode indices as
$(x,y,\uparrow)\mapsto 6x+2y$ and
$(x,y,\downarrow)\mapsto 6x+2y+1$.
Blue and red bonds describe vertical hopping,
green bonds describe horizontal hopping,
and orange dotted links represent on-site
density--density interactions between opposite spins.
}
\label{fig:FH23}
\end{figure}

After estimating the time-domain correlator $C(A,B,t)$ on a discrete time grid, we reconstruct its frequency content using MUSIC, which builds a Hankel matrix from the sampled signal and decomposes it into a signal subspace and an orthogonal noise subspace. From the noise subspace one forms the MUSIC noise-subspace correlation function: 
\begin{equation}
R(\omega) \coloneqq \bigl\|U_{\mathrm{noise}}^\dagger a(\omega)\bigr\|_2,
\end{equation}
where $a(\omega)=[e^{-\rmi \omega n}]_{n=1}^{N}$ and $U_{\mathrm{noise}}$ spans the noise subspace; see \aref{app:signalanalysis} for more details. In the ideal noiseless case and with sufficient sampling, $U_{\mathrm{noise}}^\dagger a(\omega)=0$ at the true frequencies. As plotted in our numerical results, $R(\omega)$ therefore provides a direct and visually stable diagnostic of the recovered spectral support and its robustness under imperfections. Compared with directly Fourier transforming finite-time data, the MUSIC noise-subspace correlation function can resolve closely spaced frequencies beyond the naive Fourier limit when the signal is well-approximated by a sparse sum of exponentials.

Once the dominant frequencies $\{\hat f_k\}$ are identified from the local minima of $R(\omega)$, one can estimate the corresponding amplitudes $\{\hat c_k\}$ by a linear least-squares fit of the model $s_n \approx \sum_{k=1}^{\chi} c_k e^{-\rmi f_k n}$ to the measured time samples. Finally, the many-body Green's function can be reconstructed by substituting the recovered frequencies and weights into the Lehmann form, i.e., as a sum of simple poles at the transition frequencies with residues determined by the fitted amplitudes.

\subsection{Ground states}

\begin{figure}[h!]
    \centering
    \includegraphics[width=0.6\columnwidth]{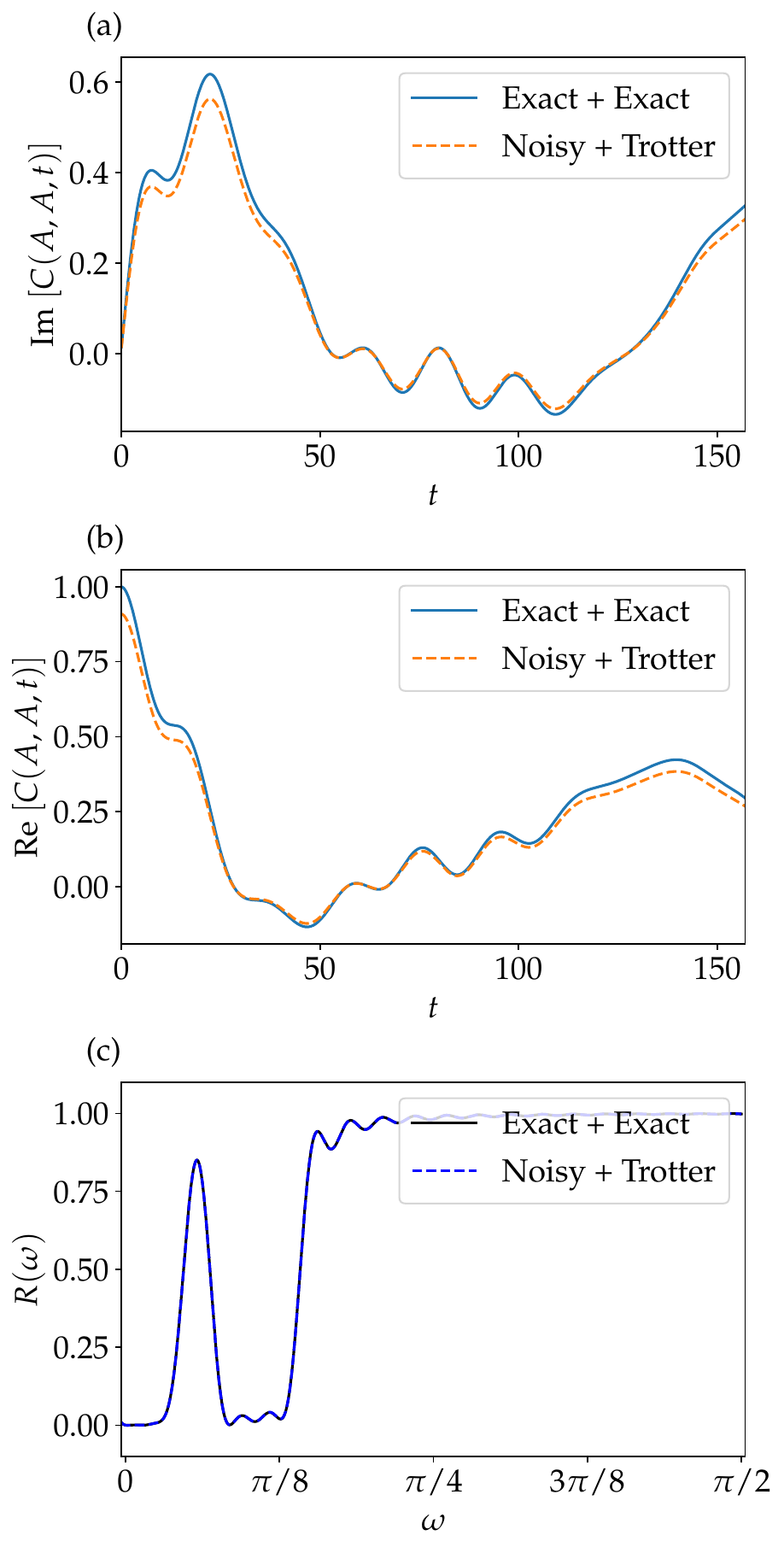}
    \caption{Ground-state numerical simulations for the 2D Fermi--Hubbard model on a $2\times 3$ lattice with open boundary conditions. The interaction strength is $h_U = 6$. The observable is $A = B = c_0 + c_0^\dag$. The system is evolved up to time $T = 200\pi$ with $N_T = 4\times 10^3$, so that the Trotter step is $\pi/20$ (only the region $t\in [0,50\pi]$ is shown). Initial-state preparation error is modeled by perturbing the initial state once as $\widetilde{\rho}=(\rho+\varepsilon\varrho)/\mathrm{tr}(\rho+\varepsilon\varrho)$ with $\varepsilon = 0.1$ and a fixed random state $\varrho$. In the legend, the first ``Exact'' denotes an exact initial state, the second ``Exact'' denotes exact time evolution; ``Noisy'' corresponds to using the perturbed initial state $\widetilde{\rho}$; and ``Trotter'' refers to simulation via the TS decomposition. (a) Imaginary part of $C(A,B,t)$ as a function of time. (b) Real part of $C(A,B,t)$. (c) Noise-subspace correlation function $R(\omega)$ for $\omega \in [0,\pi/2]$, constructed from the sampled time signal.} 
    \label{fig:ground_strong}
\end{figure}

\begin{figure}[h!]
    \centering
    \includegraphics[width=0.6\columnwidth]{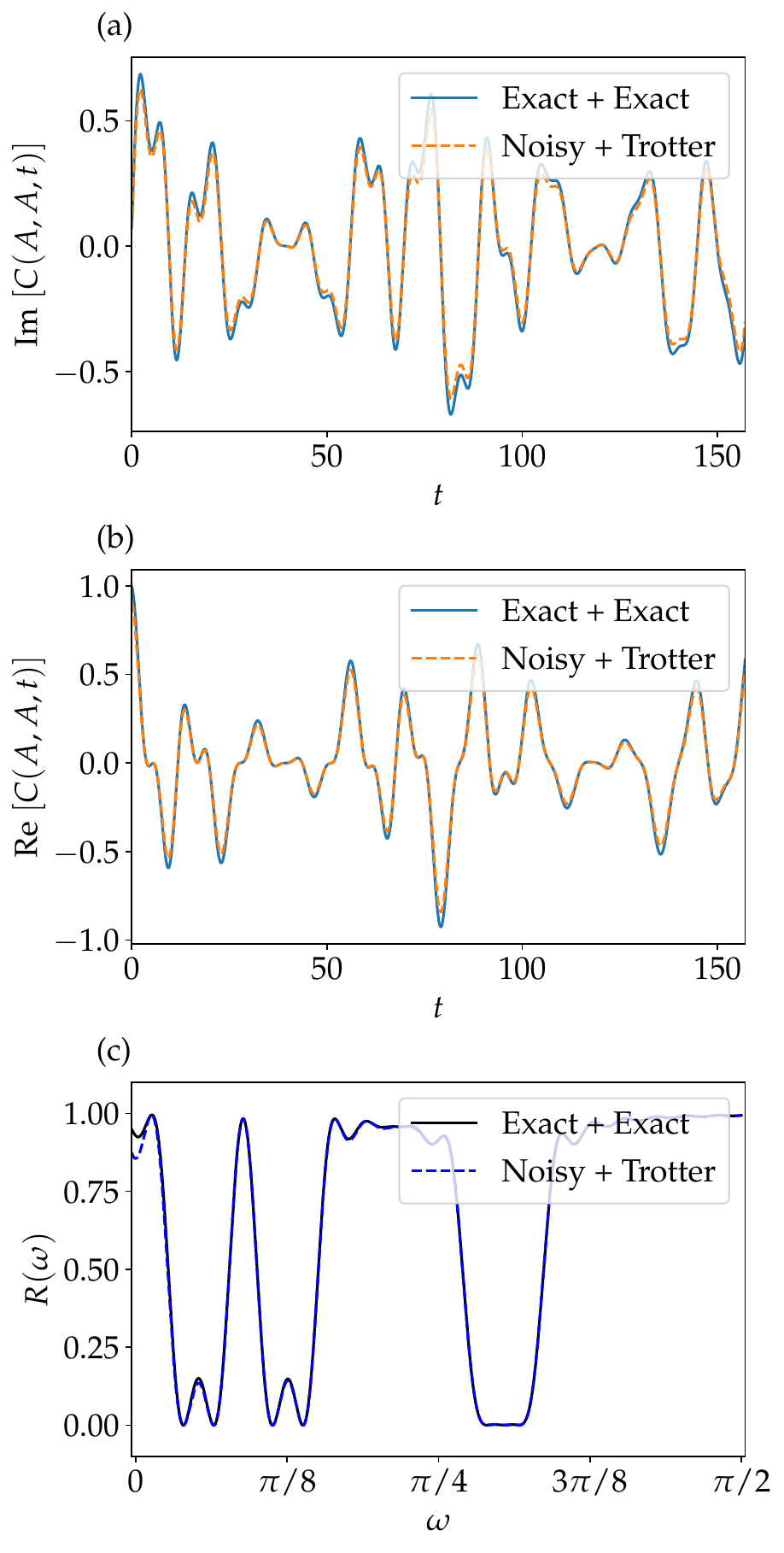}
    \caption{Ground-state numerical simulations for the 2D Fermi--Hubbard model on a $2\times 3$ lattice with open boundary conditions. The setup and the meaning of the subplots are the same as \fref{fig:ground_strong}, except the interaction strength is $h_U = 0.1$.}
    \label{fig:ground_weak}
\end{figure}

In \fref{fig:ground_strong} and \fref{fig:ground_weak} we evaluate the ground-state time correlator $C(A,B,t)$ that underlies the corresponding Green’s function and benchmark against both exact evolution and first-order TS simulation. We consider two interaction regimes: a strongly correlated case ($h_U=6$) and a weakly interacting case ($h_U=0.1$), with $A=B=c_0+c_0^\dagger$. Across both regimes, the reconstructed real and imaginary parts of the correlator remain stable under moderate initial-state preparation error and under Trotterization at the chosen step size. Moreover, the noise-subspace correlation function $R(\omega)$ computed from the noisy data closely matches its exact counterpart, indicating that the signal-processing stage is robust to the state-preparation perturbations and Trotter errors considered here.

\subsection{Finite-temperature thermal states}

\begin{figure}[h!]
    \centering
    \includegraphics[width=0.6\columnwidth]{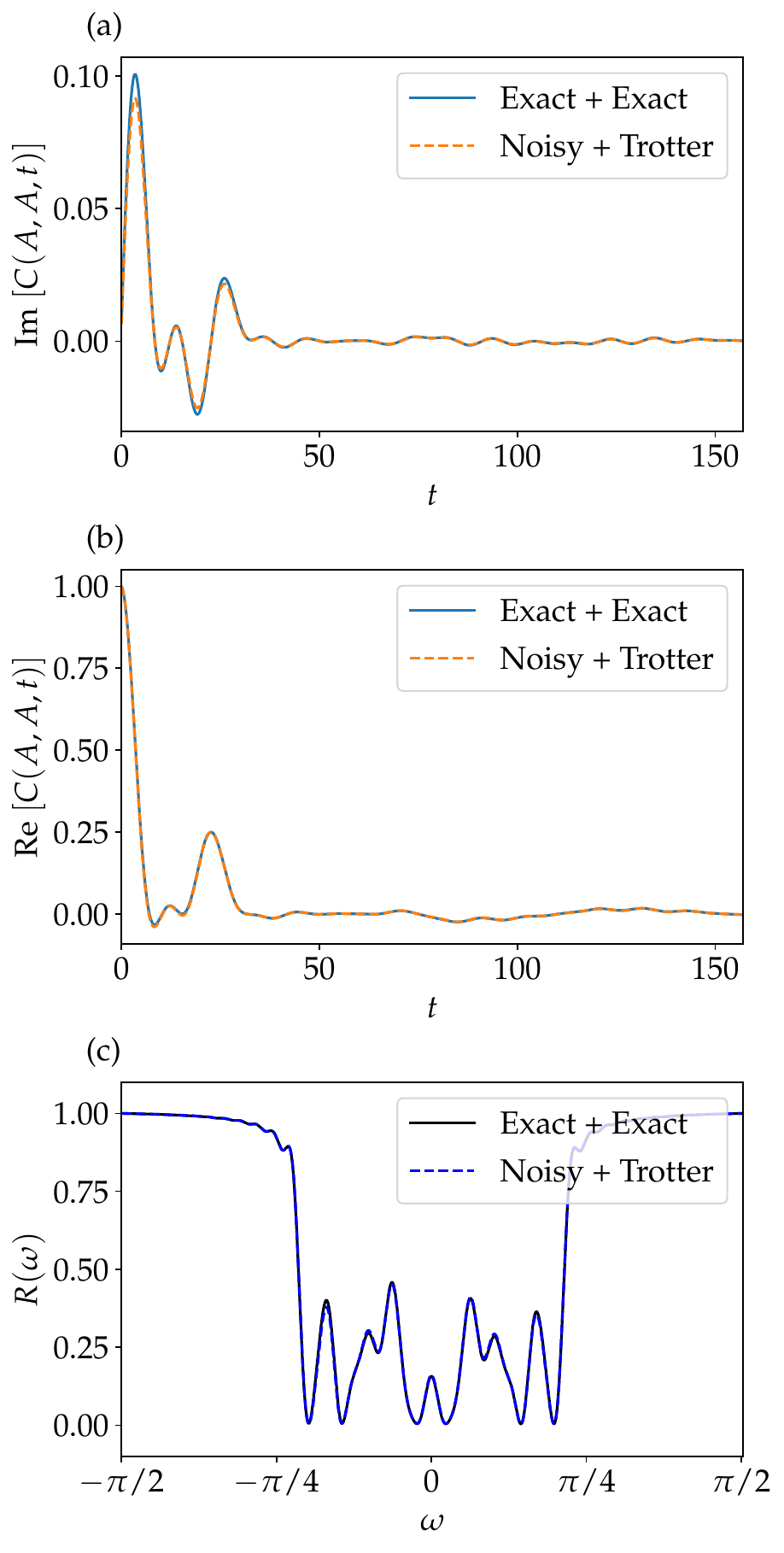}
    \caption{Finite-temperature numerical simulations for the 2D Fermi--Hubbard model on a $2\times 3$ lattice with open boundary conditions. The setup and the meaning of the subplots are the same as \fref{fig:ground_strong}, except the initial state is the thermal state $\rho_\beta$ with $\beta = 1$ and the interaction strength is $h_U = 6$. }
    \label{fig:thermal_strong}
\end{figure}

\begin{figure}[h!]
    \centering
    \includegraphics[width=0.6\columnwidth]{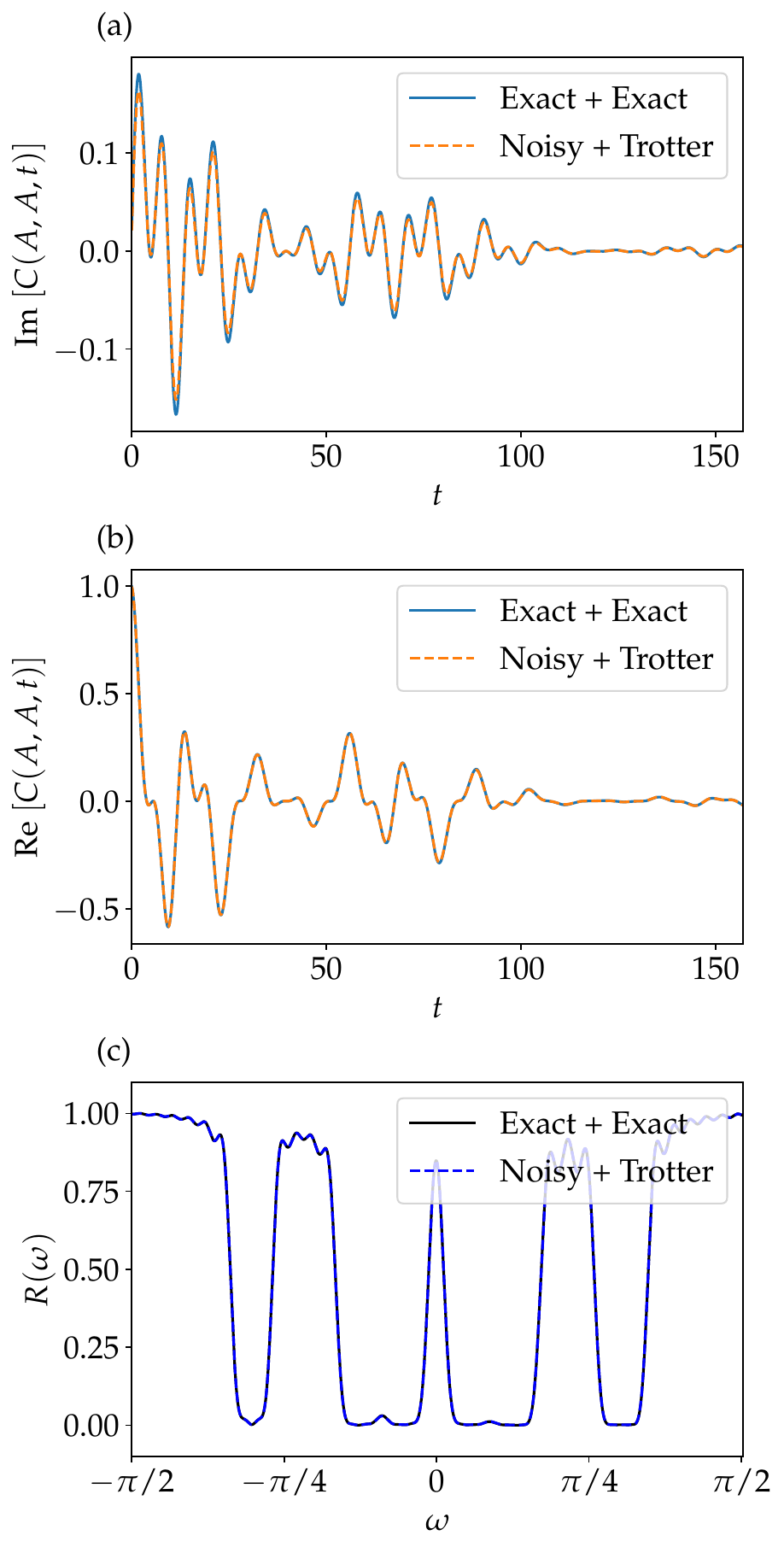}
    \caption{Finite-temperature numerical simulations for the 2D Fermi--Hubbard model on a $2\times 3$ lattice with open boundary conditions. The setup and the meaning of the subplots are the same as \fref{fig:ground_strong}, except the interaction strength is $h_U = 0.1$.}
    \label{fig:thermal_weak}
\end{figure}

Figures \ref{fig:thermal_strong} and \ref{fig:thermal_weak} extend the same analysis to finite temperature with $\beta=1$. In contrast to the ground-state setting, the correlator receives contributions from multiple energy differences weighted by thermal populations, which generally increases spectral congestion. Nevertheless, the TQS estimator continues to yield accurate real and imaginary components of $C(A,A,t)$ in both interaction regimes. These results support the practicality of the symmetry-enabled protocol for probing equilibrium dynamics beyond zero temperature, provided that the parity symmetry condition is satisfied and the thermal-state preparation error remains controlled.

\subsection{Finite-temperature OTOCs}

\begin{figure}[h!]
    \centering
    \includegraphics[width=0.6\columnwidth]{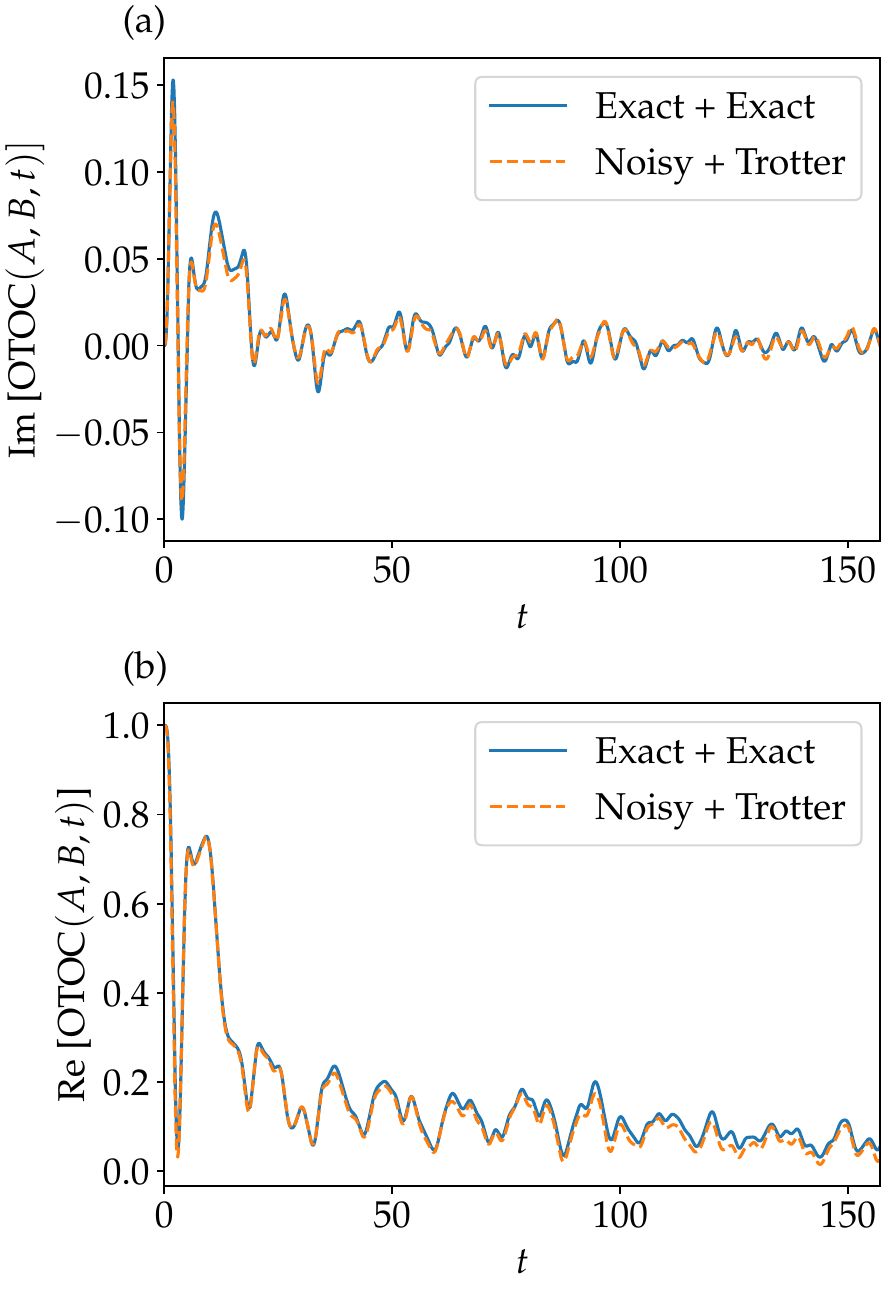}
    \caption{Real and imaginary components of the OTOC for the 1D XXZ model [see \eref{eq:XXZ}] on $N = 8$ sites with open boundary conditions. The parameters are $J_X = 1$, $J_Z = 2$, and $h = 1$. The local observables are $A = X_0$ and $B = Z_2$. The initial state is the thermal state $\rho_\beta$ with $\beta = 1$. The remaining setup and the meaning of the subplots are similar to those in \fref{fig:ground_strong}.}
    \label{fig:otoc_xxz}
\end{figure}

Figure~\ref{fig:otoc_xxz} illustrates the extension of TQS to out-of-time-order correlators (OTOCs) for the one-dimensional XXZ chain defined in \eref{eq:XXZ}. We set $J_X = 1$, $J_Z = 2$, and $h = 1$, corresponding to an anisotropy parameter $\Delta = J_Z/J_X = 2 > 1$ in the presence of a uniform longitudinal field. At zero field, a chain with $\Delta > 1$ lies in the gapped Néel phase. A longitudinal field exceeding the lower critical field $h_{c1}$, however, closes the gap and drives the system into a partially magnetized, gapless Luttinger liquid phase, which persists up to the saturation field $h_{c2}$~\cite{Schulz1986}. For the parameters above, $h$ lies well within this window ($h_{c1} < h < h_{c2}$). Because the longitudinal field commutes with the total magnetization $\sum_i S^z_i$, the model remains Bethe-ansatz integrable. Its dynamics are nevertheless genuinely interacting and nontrivial.

This model is chosen because gapless systems are generally demanding to simulate classically. Their nonequilibrium dynamics generate rapidly growing entanglement, which limits tensor-network methods. This makes the model a suitable benchmark for demonstrating the potential advantage of quantum hardware should the protocol be implemented experimentally. It also provides a representative setting in which to assess the robustness of OTOC estimation against Trotterization errors and imperfections in state preparation.

\section{Conclusion and outlook}
\label{sec:conclude}

We introduced a symmetry-enabled direct quantum protocol, tailored quench spectroscopy (TQS), for computing both the real and imaginary parts of two-point time correlators without controlled unitaries or ancilla qubits in the correlator-estimation circuit. The central observation is that when a parity operator $P$ commutes with the Hamiltonian and the observables of interest are parity-odd, symmetry enforces selection rules that simplify quench spectroscopy readout. For fixed-parity initial states (including energy eigenstates) the correlator can be obtained from only two quench functions, and for thermal states we showed how to combine parity-resolved correlators obtained from symmetric and antisymmetric Gibbs states. Together with classical signal processing (e.g., MUSIC), these correlators yield the Lehmann representation of the Green's function. Our error analysis highlights that symmetry-based symmetrization can quadratically suppress parity-breaking state-preparation errors. Numerical simulations on Fermi--Hubbard and XXZ instances support the robustness of the approach under Trotterization and imperfect state preparation. Finally, we demonstrated that the same symmetry-based measurement primitive extends to a class of multi-point correlators, including OTOCs.

This work opens several promising directions. First, it is natural to ask how far symmetry-enabled direct protocols can be extended beyond linear two-point response. One avenue is the measurement of higher-order and nonlinear response functions \cite{ono2025extracting}, where parity constraints might reduce the overhead of accessing specific operator orderings or nested commutators. In terms of the symmetric thermal states, other Lindbladians, such as \cite{chen2025efficient,ding2025efficient}, can also be adapted. A related question is whether analogous ideas can be adapted to non-Hermitian \cite{pan2020non,geier2022from} or effective open-system settings, e.g., when the dynamics is generated by a non-Hermitian Hamiltonian or by a Lindbladian, where Green's functions and response theory take modified forms and symmetry constraints may still yield simplifications.

Second, a near-term priority is experimental demonstration. Because TQS requires only native time evolution under a time-independent Hamiltonian and end-of-circuit measurements, it is compatible with both digital platforms and analog simulators, where quench-style protocols are natural. A promising first milestone is a hardware demonstration of parity-assisted recovery of $\Re[C(A,B,t)]$ and $\Im[C(A,B,t)]$ in small Hubbard or spin-chain instances, followed by reconstruction of low-lying spectral features via classical signal analysis. Benchmarking the protocol under realistic noise and comparing against other available approaches would clarify the practical advantage of symmetry-enabled direct measurement.

Finally, several technical questions remain open. On the theory side, sharper sample-complexity guarantees for signal reconstruction under realistic noise (including state-preparation and measurement noise) would strengthen the end-to-end performance picture. On the implementation side, developing more efficient and hardware-tailored preparation methods for parity-resolved thermal states, possibly avoiding dissipative overhead in some settings, would broaden the applicability of finite-temperature TQS. We expect that exploring these directions will further establish symmetry-enabled direct protocols as practical tools for probing finite-temperature dynamics of strongly correlated quantum matter.

\section*{Acknowledgements}

C.Y.~acknowledges support from the National Key Research and Development Program of China (Grant No.~2022YFA1404204),  Shanghai Municipal Science and Technology Major Project (Grant No.~2019SHZDZX01), National Natural Science Foundation of China (Grant No.~92165109), and Shanghai Municipal Science and Technology Major Project (Grant No. 2019SHZDZX01-ZX04). C.Z. acknowledges support from the Natural Sciences and Engineering Research Council of Canada (NSERC) through Discovery Grant RGPIN-2026-06304; from the Faculté des sciences and the Institut quantique at Université de Sherbrooke; and from the Institut transdisciplinaire d’information quantique (INTRIQ), a strategic cluster funded by the Fonds de recherche du Québec – Nature et technologies. We sincerely thank Dr.~Mario Motta and Dr.~Alexandre Foley for their careful reading of our draft and for the numerous insightful comments they provided.

\section*{Author contributions}

Both C.Y. and C.Z. conceived the core idea, discussed the results, and contributed to writing the manuscript. C.Y. performed the numerical simulations. ChatGPT was used to improve the clarity of the writing and assist in identifying mathematical errors.

\section*{Data availability}
The data and code supporting the findings of this work are available at~\cite{tqs2025}.

\clearpage

\let\oldaddcontentsline\addcontentsline
\renewcommand{\addcontentsline}[3]{}

\bibliographystyle{plainnat}
\bibliography{main}

\let\addcontentsline\oldaddcontentsline

\newpage
\appendix

\setcounter{equation}{0}
\setcounter{figure}{0}
\setcounter{table}{0}
\setcounter{theorem}{0}
\setcounter{proposition}{0}
\setcounter{lemma}{0}
\setcounter{section}{0}
\setcounter{page}{1}

\section{Supplementary background}

\subsection{Many-Body Green's Function}
\label{app:background-green}
In this appendix we give an overview of basic facts about many-body Green's function in condensed matter physics. Suppose $H$ is a Hamiltonian with a non-degenerate ground state and has composition $H = \sum_{n=0}^{d-1}E_n |n\>\<n|$. Let $O_a,O_b$ be two linear operators. The many-body Green's function at zero-temperature is defined as:
\begin{equation}
    G(z) \coloneqq \langle0|O_a (z - H + E_0)^{-1}O_b|0\rangle = \sum_n \frac{\< 0|O_a|n\>\< n|O_b|0\>}{z - E_n + E_0},\  \Im(z) > 0.
\end{equation}
 Note that for all $\zeta$ with $\Re(\zeta) > 0$, we have $\int_0^{\infty} e^{-\zeta t}dt = \zeta^{-1}$.
Eigenstate decomposition yields
\begin{equation}
    C(O_a,O_b,-t) = \sum_n \langle 0| O_a|n\rangle\langle n|e^{-\rmi Ht}O_b e^{\rmi Ht}|0\rangle = \sum_n \langle 0|O_a|n\rangle\langle n|O_b|0\rangle e^{\rmi(E_0 - E_n)t}.
\end{equation}
Note that for all $z$ with positive imaginary component, we have
\begin{gather}
    \int_0^{\infty} e^{\rmi zt} e^{\rmi(E_0 - E_n)t}dt = \frac{\rmi}{z - E_n + E_0}.
\end{gather}
Accordingly, 
\begin{equation}
    G(z) = -\rmi \int_0^{\infty}e^{\rmi zt}C(O_a,O_b,-t)dt.
\end{equation}
This confirms \eref{eq:laplace} in the main text. The purpose of this transformation is to express the Green's function, which is defined in terms of the resolvent \((z-H+E_0)^{-1}\), as the Laplace/Fourier transform of a real-time correlation function. This representation is useful because the time-dependent correlation function can be obtained directly from the real-time evolution generated by \(H\), providing a practical route to estimating the Green's function without explicitly implementing the resolvent.

Let $\Delta_n=E_n-E_0$ and define
\begin{equation}
    w_n
    =
    \langle 0|O_a|n\rangle
    \langle n|O_b|0\rangle.
\end{equation}
In general, $w_n$ may be complex.
For $z=\omega+\rmi\eta$, with
$\omega\in\bbR$ and $\eta>0$, we have
\begin{equation}
    G(\omega+\rmi\eta)
    =
    \sum_n
    w_n
    \frac{\omega-\Delta_n-\rmi\eta}
    {(\omega-\Delta_n)^2+\eta^2}.
\end{equation}

Define the broadened spectral function
\begin{equation}
    S(\omega,\eta)
    \coloneqq
    -2\operatorname{Im}G(\omega+\rmi\eta).
\end{equation}
Taking the imaginary part gives
\begin{equation}
    S(\omega,\eta)
    =
    2\sum_n
    \frac{
        \eta\operatorname{Re}w_n
        -(\omega-\Delta_n)\operatorname{Im}w_n
    }{
        (\omega-\Delta_n)^2+\eta^2
    }.
\end{equation}

For any $z$ with $\operatorname{Im}z>0$,
the Cauchy transform of $S$ satisfies
\begin{align}
    \int_{-\infty}^{\infty}
    \frac{d\omega}{2\pi}
    \frac{S(\omega,\eta)}{z-\omega}
    =
    \sum_n
    \frac{w_n}{z-\Delta_n+\rmi\eta}
    =G(z+\rmi\eta).
\end{align}
Consequently, the Green's function is
recovered in the limit
\begin{equation}
    G(z)
    =
    \lim_{\eta\to0^+}
    \int_{-\infty}^{\infty}
    \frac{d\omega}{2\pi}
    \frac{S(\omega,\eta)}{z-\omega},
    \qquad \operatorname{Im}z>0.
\end{equation}

\subsection{Jordan--Wigner Transformation}
\label{app:jordan-wigner}
Jordan--Wigner transformation maps fermionic creation and annihilation operators to Pauli strings on qubits. 
For a fermionic lattice with $N$ sites and two spin species, there are $J=2N$ spin-orbitals, which we index by $j\in\{0,1,\dots,J-1\}$. 
The standard Jordan--Wigner mapping is
\begin{equation}
    c_j=\Big(\prod_{n=0}^{j-1}Z_n\Big)\sigma_j^-,
    \qquad 
    \sigma_j^\pm=\frac{X_j\pm \rmi Y_j}{2}.
\end{equation}
From this mapping one obtains the identities
\begin{equation}
    c_j c_j^\dagger=\frac{1-Z_j}{2},
    \qquad
    n_j\coloneqq c_j^\dagger c_j=\frac{1+Z_j}{2},
\end{equation}
and, for nearest neighbors,
\begin{equation}
    c_j^\dagger c_{j+1}=-\sigma_j^+\sigma_{j+1}^-,
    \qquad
    c_{j+1}^\dagger c_j=-\sigma_{j+1}^+\sigma_j^-,
\end{equation}
which imply the familiar hopping relation
\begin{equation}
    c_j^\dagger c_{j+1}+c_{j+1}^\dagger c_j
    =-\frac{X_jX_{j+1}+Y_jY_{j+1}}{2}.
\end{equation}
Finally, note that fermionic bilinears do not commute in general: for example,
\begin{equation}
    [c_i^\dagger c_j,\;c_j^\dagger c_k]=c_i^\dagger c_k\neq 0
\end{equation}
whenever the indices overlap.

\subsection{Davies generators and Gibbs-state preparation}
\label{app:davies}

We briefly review the Davies generator, a canonical Markovian open-system dynamics describing the weak coupling of a quantum system to a thermal bath. Under the usual weak-coupling, Markovian, and secular approximations, the reduced dynamics of the system is generated by a Lindbladian whose jump operators resolve the energy differences of the system Hamiltonian. When the transition rates satisfy the thermal Kubo--Martin--Schwinger (KMS) condition, the Gibbs state is stationary under the resulting dynamics.

Let
\begin{equation}
    H=\sum_n E_n |\psi_n\rangle\langle\psi_n|
\end{equation}
be the system Hamiltonian. For an arbitrary linear operator $J$, define its Heisenberg evolution
    $J(t)=e^{\rmi Ht}J e^{-\rmi Ht}$.
The operator can be decomposed into its Bohr-frequency components as
\begin{equation}
    J(t)=\sum_{\nu\in\mathcal D(H)}J_\nu e^{\rmi\nu t},
\end{equation}
where
\begin{equation}
    \mathcal D(H)
    :=
    \{E_m-E_n:\,m,n\}
\end{equation}
is the set of Bohr frequencies and
\begin{equation}
    J_\nu
    =
    \sum_{m,n:\,E_m-E_n=\nu}
    \langle\psi_m|J|\psi_n\rangle
    |\psi_m\rangle\langle\psi_n|.
\end{equation}
In particular, when $J$ is Hermitian, $J_\nu^\dagger=J_{-\nu}$, these components satisfy
\begin{equation}
    [H,J_\nu]=\nu J_\nu,
    \qquad
    (J_\nu)^\dagger=(J^\dagger)_{-\nu}.
\end{equation}
Equivalently, $J_\nu$ maps an energy eigenspace of energy $E$ into one of
energy $E+\nu$.

For simplicity, consider a single Hermitian system coupling operator $J$.
Given nonnegative transition rates $\eta(\nu)\ge 0$, the dissipative part of
the Davies generator takes the form
\begin{equation}
    \mathcal L[\rho]
    =
    \sum_{\nu\in\mathcal D(H)}
    \eta(\nu)
    \left(
        J_\nu\rho J_\nu^\dagger
        -\frac{1}{2}
        \{J_\nu^\dagger J_\nu,\rho\}
    \right).
\end{equation}
More generally, one may introduce a family of coupling operators
$\{J_a\}_a$ and corresponding rates $\eta_a(\nu)$, in which case the
generator is obtained by summing over both $a$ and $\nu$. The full Davies
generator may also contain a Lamb-shift Hamiltonian term
$-\rmi[H_{\rm LS},\rho]$, where $H_{\rm LS}$ commutes with $H$; since this
term does not affect the thermal fixed-point structure discussed below, we
omit it here.

The adjoint generator in the Heisenberg picture acts as
\begin{equation}
    \mathcal L^\dagger[O]
    =
    \sum_{\nu\in\mathcal D(H)}
    \eta(\nu)
    \left(
        J_\nu^\dagger OJ_\nu
        -\frac{1}{2}
        \{J_\nu^\dagger J_\nu,O\}
    \right).
\end{equation}
By construction, the Davies generator is covariant with respect to the
Hamiltonian evolution: its dissipative terms do not mix different Bohr
frequencies. This frequency resolution is a consequence of the secular
approximation and is one of the characteristic structural properties of
Davies dynamics.

Let $\rho_\beta$
denote the Gibbs state at inverse temperature $\beta$. The Bohr-frequency
operators obey the modular relations
\begin{equation}
    \rho_\beta J_\nu
    =
    e^{-\beta\nu}J_\nu\rho_\beta,
    \qquad
    \rho_\beta^{1/2}J_\nu
    =
    e^{-\beta\nu/2}J_\nu\rho_\beta^{1/2}.
\end{equation}
Thermal equilibrium of the bath imposes the KMS relation on the transition
rates,
\begin{equation}
    \eta(-\nu)
    =
    e^{\beta\nu}\eta(\nu),
    \qquad
    \nu\in\mathcal D(H).
\end{equation}
Thus transitions that increase and decrease the system energy occur with
relative weights determined by the Boltzmann factor.

Under this condition, the Davies generator satisfies KMS detailed balance
with respect to $\rho_\beta$. In the present convention, this can be written
as
\begin{equation}
    \mathcal L[X]
    =
    \rho_\beta^{1/2}
    \mathcal L^\dagger
    \left[
        \rho_\beta^{-1/2}
        X
        \rho_\beta^{-1/2}
    \right]
    \rho_\beta^{1/2}.
\end{equation}
In particular,
\begin{equation}
    \mathcal L[\rho_\beta]=0,
\end{equation}
so the Gibbs state is a stationary state of the dynamics. If the jump
operators connect all relevant energy and symmetry sectors, so that the
Davies semigroup is primitive, $\rho_\beta$ is moreover the unique
full-rank stationary state. Without such an ergodicity condition, additional
stationary states may exist.

The Davies generator therefore provides a canonical connection between
thermal equilibrium and Markovian quantum dynamics: its jump operators are
obtained by resolving physical system--bath couplings into the Bohr
frequencies of $H$, while the KMS relation fixes the relative rates of
opposite-frequency transitions and guarantees thermal stationarity.

\section{Proof of Main Results}
\label{app:proofmainresults}

\subsection{Proof of \lref{lem:parity}}
\label{sec:proof2}
\begin{proof}
Observe that 
\begin{equation}
    \<n|A|m\> = -\<n|PAP|m\> = -p_mp_n\<n|A|m\>. 
\end{equation}
Hence, if $p_np_m = 1$, then $\<n|A|m\> = 0$. Conversely, if $\<n|A|m\> \neq 0$, then $p_np_m = -1$. This completes the proof of \lref{lem:parity}.
\end{proof}

\subsection{Proof of \thref{thm:eigenstate}}
\label{sec:proof3}

\begin{lemma} \label{lem:commute}
    Suppose $\rho$ commutes with $P$, and observables $O_1,\ldots,O_L$ all anti-commute with $P$. If $L$ is odd, then
    \begin{equation}
        \mathrm{tr}(\rho \prod_{\ell=1}^L O_\ell) = 0.
    \end{equation}
\end{lemma}
\begin{proof}
    Note that
    \begin{equation}
        P\prod_{\ell=1}^L O_\ell = -\prod_{\ell=1}^L O_\ell P.
    \end{equation}
    Hence,
    \begin{equation}
        \mathrm{tr}(\rho \prod_{\ell=1}^L O_\ell) = \mathrm{tr}(P\rho P \prod_{\ell=1}^L O_\ell) = -\mathrm{tr}(\rho \prod_{\ell=1}^L O_\ell P^2) = -\mathrm{tr}(\rho \prod_{\ell=1}^L O_\ell).
    \end{equation}
\end{proof}

With the above observation, we present the proof of \thref{thm:eigenstate} as follows.
\begin{proof}
By definition, we have
\begin{equation}
    Q_{\rho,B}\left(e^{-\rmi Ht}A\right) = \mathrm{tr}(e^{-\rmi Ht}A\rho A e^{\rmi Ht} B) = \mathrm{tr}(\rho A B(t)A).
\end{equation}
Since $\{A,P\} = \{B(t),P\} = [\rho,P] = 0$, in conjunction with \lref{lem:commute}, we have $Q_{\rho,B}\left(e^{-\rmi Ht}A\right)= 0$. Similarly, we have $Q_{\rho,B}\left(e^{-\rmi Ht}\right) = Q_{\rho,B}\left(e^{-\rmi Ht}P\right) = 0$. This confirms that
\begin{align}
    Q_{\rho,B}\left(e^{-\rmi Ht}U_{\mathrm{Im}}\right) &= \mathrm{tr}\left[e^{-\rmi Ht}\frac{I + \rmi A}{\sqrt{2}}\rho \frac{I - \rmi A}{\sqrt{2}} e^{\rmi Ht} B\right] \nonumber \\
    &= \Im[C(A,B,t)] + \frac{1}{2}Q_{\rho,B}(e^{-\rmi Ht}) + \frac{1}{2}Q_{\rho,B}(e^{-\rmi Ht}A) \nonumber \\
    &= \Im[C(A,B,t)], \\
    Q_{\rho,B}\left(e^{-\rmi Ht}U_{\mathrm{Re}}\right) &= \mathrm{tr}\left[e^{-\rmi Ht}\frac{P + A}{\sqrt{2}}\rho \frac{P + A}{\sqrt{2}} e^{\rmi Ht} B\right] \nonumber \\
    &= \Re[C(PA,B,t)] + \frac{1}{2}Q_{\rho,B}(e^{-\rmi Ht}P) + \frac{1}{2}Q_{\rho,B}(e^{-\rmi Ht}A) \nonumber\\
    &= \Re[C(PA,B,t)].
\end{align}
Because $\rho P = p\rho$ with $p \in \{\pm 1\}$, we further have
    \begin{equation}
        C(PA,B,t) = \mathrm{tr}(\rho PA B(t)) = p\mathrm{tr}(\rho AB(t)) = pC(A,B,t).
    \end{equation} This completes the proof of \thref{thm:eigenstate}.
\end{proof}

\subsection{Proof of \thref{thm:thermalstate}}
\label{app:proofofthm2}

\begin{proof}
Recall the decomposition
\begin{equation}
\rho_\beta
= \frac{1 + \mathrm{tr}(\rho_\beta P)}{2}\,\rho_\mathrm{S}
+ \frac{1 - \mathrm{tr}(\rho_\beta P)}{2}\,\rho_\mathrm{A},
\end{equation}
where $P\rho_\mathrm{S}=\rho_\mathrm{S}$ and $P\rho_\mathrm{A}=-\rho_\mathrm{A}$.
By linearity of the correlator in the state, we have
\begin{equation}
C(A,B,t)=\mathrm{tr}(\rho_\beta A B(t))
=\frac{1 + \mathrm{tr}(\rho_\beta P)}{2}\mathrm{tr}(\rho_\mathrm{S} A B(t))
+\frac{1 - \mathrm{tr}(\rho_\beta P)}{2}\mathrm{tr}(\rho_\mathrm{A} A B(t)).
\end{equation}
Taking real parts yields
\begin{equation}
\Re\left[C(A,B,t)\right]
=\frac{1 + \mathrm{tr}(\rho_\beta P)}{2}\Re\left[\mathrm{tr}(\rho_\mathrm{S} A B(t))\right]
+\frac{1 - \mathrm{tr}(\rho_\beta P)}{2}\Re\left[\mathrm{tr}(\rho_\mathrm{A} A B(t))\right].
\end{equation}
Now apply \thref{thm:eigenstate} to each parity sector. Since $\rho_\mathrm{S}$ has definite parity $p=+1$ and $\rho_\mathrm{A}$ has definite parity $p=-1$, we obtain
\begin{equation}
\Re\left[\mathrm{tr}(\rho_\mathrm{S} A B(t))\right]
= Q_{\rho_\mathrm{S},B}\!\left(e^{-\rmi Ht}U_{\mathrm{Re}}\right),
\quad
\Re\left[\mathrm{tr}(\rho_\mathrm{A} A B(t))\right]
= -\,Q_{\rho_\mathrm{A},B}\!\left(e^{-\rmi Ht}U_{\mathrm{Re}}\right).
\end{equation}
Substituting into the previous expression gives \eref{eq: theorem2}, completing the proof.
\end{proof}

\section{Signal analysis algorithm}
\label{app:signalanalysis}

We use the MUltiple SIgnal Classification (MUSIC)
algorithm \cite{schmidt86multiple,liao2016music}
to reconstruct the frequencies appearing in
time-domain correlators.

Consider a signal
\begin{equation}\label{eq:signal}
    s_n=\sum_{k=1}^{\chi}c_k e^{-\rmi f_k n},
    \qquad n=0,\ldots,2N-2,
\end{equation}
where $f_k\in[0,2\pi)$ are distinct frequencies
modulo $2\pi$, $c_k\in\mathbb C\setminus\{0\}$,
and $\chi<N$ is the number of spectral lines.

Construct the Hankel matrix $M\in\mathbb C^{N\times N}$,
\begin{equation}
    M_{m,n}=s_{m+n},
    \qquad m,n=0,\ldots,N-1.
\end{equation}
Writing
\begin{equation}
    V_{m,k}=e^{-\rmi f_k m},
    \qquad
    D=\operatorname{diag}(c_1,\ldots,c_\chi),
\end{equation}
we obtain the Vandermonde factorization
\begin{equation}
    M=VDV^{\mathsf T},
    \qquad \operatorname{rank}(M)=\chi.
\end{equation}

Let $M=U\Sigma V_M^\dagger$ be its singular
value decomposition, with singular values
in non-increasing order. Partition
\begin{equation}
    U=[U_{\rm sig}\ \ U_{\rm noise}],
\end{equation}
where $U_{\rm noise}\in
\mathbb C^{N\times(N-\chi)}$ spans
the orthogonal complement of the signal subspace.

Define the normalized steering vector
\begin{equation}
    a(\omega)=
    \frac{1}{\sqrt N}
    [e^{-\rmi\omega n}]_{n=0}^{N-1}.
\end{equation}
The MUSIC noise-space correlation function is
\begin{equation}\label{eq:signalnoise}
    R(\omega)=
    \|U_{\rm noise}^\dagger a(\omega)\|_2.
\end{equation}
In the noiseless case,
\begin{equation}
    R(\omega)=0
    \quad\Longleftrightarrow\quad
    \omega\in\{f_1,\ldots,f_\chi\}
    \pmod{2\pi}.
\end{equation}

We next recall a deterministic stability bound
from Ref.~\cite{liao2016music}.

\begin{theorem}[MUSIC subspace stability]
\label{thm:music_stability}
Let $\widetilde M=M+E$ be the Hankel matrix
constructed from noisy signal samples.
Denote the nonzero singular values of $M$ by
\begin{equation}
    \sigma_1(M)\ge\cdots\ge\sigma_\chi(M)>0.
\end{equation}
Suppose
\begin{equation}
    \epsilon_H\coloneqq\|E\|_2
    <\sigma_\chi(M).
\end{equation}
Let $\widetilde R(\omega)$ be the MUSIC
noise-space correlation function obtained
from the leading $\chi$ left singular vectors
of $\widetilde M$. Then
\begin{equation}\label{eq:MUSIC}
    \sup_{\omega\in[0,2\pi)}
    |\widetilde R(\omega)-R(\omega)|
    \le
    \frac{
       4\sigma_1(M)+2\epsilon_H
    }{
       [\sigma_\chi(M)-\epsilon_H]^2
    }\epsilon_H.
\end{equation}
\end{theorem}

The bound is deterministic and applies to arbitrary
additive errors. Its stability depends on the
smallest nonzero singular value of the noiseless
Hankel matrix.

For independent, centered sub-Gaussian sample
errors with scale parameter $\sigma$, standard
random-Hankel estimates give, with high probability,
\begin{equation}
    \epsilon_H
    =O\!\left(\sigma\sqrt{N\log N}\right).
\end{equation}
If the frequencies are sufficiently separated
and the nonzero amplitudes are bounded above and
away from zero so that
\begin{equation}
    \sigma_1(M)=O(N),
    \qquad
    \sigma_\chi(M)=\Omega(N),
\end{equation}
with constants independent of $N$, then,
in the perturbative regime,
\begin{equation}
    \sup_\omega
    |\widetilde R(\omega)-R(\omega)|
    =
    O\!\left(
      \sigma\sqrt{\frac{\log N}{N}}
    \right).
\end{equation}
Sufficient quantitative separation conditions
are given in Ref.~\cite{liao2016music}.

Importantly, stability of the noise-space correlation
does not by itself imply an explicit error bound
for the estimated frequencies. Such a bound
additionally requires control of the local
behavior of $R(\omega)$ near its zeros.
Frequency-estimation bounds under further
nondegeneracy and separation assumptions are
discussed in Ref.~\cite{liao2016music}.

\begin{figure}[h!]
\begin{algorithm}[H]
{\small
\textbf{Input:}
Noisy samples $(\widetilde s_n)_{n=0}^{2N-2}$;
model order $\chi<N$.

\textbf{Output:}
Frequency estimates
$\{\widehat f_k\}_{k=1}^{\chi}$.

\begin{algorithmic}[1]
\caption{\label{alg:music}MUSIC (basic form)}
\State Construct the Hankel matrix
$\widetilde M_{m,n}=\widetilde s_{m+n}$,
$m,n=0,\ldots,N-1$.
\State Compute the SVD
$\widetilde M=\widetilde U
\widetilde\Sigma\widetilde V^\dagger$.
\State Set
$\widetilde U_{\rm noise}
=[\widetilde u_{\chi+1},\ldots,
\widetilde u_N]$.
\State Evaluate
$\widetilde R(\omega)=
\|\widetilde U_{\rm noise}^\dagger
a(\omega)\|_2$
on a frequency grid in $[0,2\pi)$.
\State Return the $\chi$ smallest local
minima, with optional local refinement.
\end{algorithmic}
}
\end{algorithm}
\end{figure}

\section{Computing time correlators via mid-circuit measurements}

In this section, we summarize a slightly simplified mid-circuit measurement (MCM) protocol (adapted from~\cite{mitarai2019methodology}) for estimating two-point correlators for general initial states. We include it here for its simplicity and generality. However, unlike our TQS protocol (which uses only end-of-circuit measurements), MCM-based estimators can be particularly sensitive to mid-circuit measurement imperfections such as readout-induced dephasing, crosstalk, and imperfect reset.

Let $A$ and $B$ be Pauli observables with eigenvalues $\pm1$. 
Measuring $A$ corresponds to the projective measurement $\{\Pi_+^A,\Pi_-^A\}$ with
\begin{equation}
    \Pi_\pm^A = \frac{I\pm A}{2}.
\end{equation}
For an initial state $\rho$, the outcome $o_1\in\{+1,-1\}$ occurs with probability
\begin{equation}
    \Pr(o_1)=\mathrm{tr}(\rho\,\Pi_{o_1}^A),
\end{equation}
and the post-measurement state is
\begin{equation}
    \rho_{o_1}=\frac{\Pi_{o_1}^A\,\rho\,\Pi_{o_1}^A}{\Pr(o_1)}.
\end{equation}
Conditioned on $o_1$, we next measure $B$ using $\Pi_\pm^B=(I\pm B)/2$, obtaining $o_2\in\{\pm1\}$. 
The joint probability is
\begin{equation}
    \Pr(o_1,o_2)
    =\mathrm{tr}\!\left(\Pi_{o_2}^B\,\Pi_{o_1}^A\,\rho\,\Pi_{o_1}^A\right).
\end{equation}
A direct calculation shows that
\begin{equation}
    \mathbb{E}[o_1o_2]
    =\sum_{o_1,o_2\in\{\pm1\}} o_1o_2\,\Pr(o_1,o_2)
    =\frac{1}{2}\mathrm{tr}\!\big(\rho(AB+BA)\big)
    =\Re\left[\mathrm{tr}(\rho AB)\right].
\end{equation}
Therefore, $\Re\left[\mathrm{tr}(\rho AB)\right]$ can be estimated by repeating the two-step measurement and averaging the product $o_1o_2$ over many shots.

\end{document}